\documentclass{nseJournal}

\usepackage[english]{babel}

\usepackage{amsmath}
\usepackage{amssymb}
\usepackage{graphicx}
\usepackage{algorithm}
\usepackage{algpseudocode}
\usepackage{amsmath}
\usepackage{setspace}
\usepackage{amssymb}
\usepackage{epsfig}
\usepackage{amscd}
\usepackage{subcaption}
\usepackage{float}
\usepackage[normalem]{ulem}

\usepackage{bm}
\usepackage{siunitx}
\usepackage[numbers]{natbib}
\usepackage{enumitem}
\usepackage[dvipsnames,svgnames,x11names]{xcolor}

\begin{document}

\title{Addressing geometrical perturbations by applying generalized polynomial chaos to virtual density in continuous energy Monte-Carlo power iteration}
\addAuthor{\correspondingAuthor{Theophile Bonnet}}{1}
\addAuthor{Anuj Dubey}{1}
\addAuthor{Eugene Shwageraus}{1}

\correspondingEmail{tlfab2@cam.ac.uk}

\addAffiliation{1}{Department of Engineering, University of Cambridge, Cambridge, UK}

\addKeyword{Monte Carlo simulation}
\addKeyword{Virtual density}
\addKeyword{Geometrical perturbations}
\addKeyword{generalized polynomial chaos}
\addKeyword{Perturbation theory}
\titlePage

\begin{abstract}
In this work, we revisit the use of the virtual density method to model uniform geometrical perturbations. We propose a general algorithm in order to estimate explicitly the effect of geometrical perturbations in continuous-energy Monte Carlo power iteration simulations. We apply the intrusive generalized polynomial chaos method in order to estimate the coefficients of a reduced model giving the multiplication factor as a function of the amplitude of the geometrical perturbation. Our method accurately estimates the reactivity change induced by uniform expansion or swelling deformations of arbitrary geometries, for a large range of deformations within a single Monte Carlo simulation. The reduced model converges rapidly in polynomial order, does not require knowledge of the adjoint flux, and is free from indirect effects. 
\end{abstract}

\section{Introduction}

In the analysis of nuclear reactors it is often necessary to look at how the system reacts to small perturbations. Amongst these perturbations are geometrical or boundary perturbations, for which the position of internal (resp. external) interfaces is perturbed. These are traditionally treated through geometry or boundary perturbation theory, respectively~\cite{larsen_boundary_1981,pomraning_boundary_1983, rahnema_boundary_1983}. Boundary perturbation theory (when the interior of the domain is not perturbed) has been extended to arbitrary order~\cite{rahnema_equivalence_1998, mckinley_order_2002}. On the other hand, the methodology introduced in~\cite{pomraning_boundary_1983} for interior boundary perturbations and complemented in~\cite{rahnema_internal_1996} has up to now been limited to first-order. For computational and algorithmic reasons, the analysis of these perturbations has long been restricted to deterministic methods, mainly due to the fact that most of such methods requires knowledge of the adjoint flux~\cite{favorite_eigenvalue_2010, favorite_revisiting_2017}. On the other hand, analysis of such geometrical perturbations using Monte Carlo simulations is not trivial, and is currently an active field of study. 

The virtual density method is an old method stating that whole-core (also called uniform) changes in geometry can be equivalently represented by turning the Boltzmann equation into its dimensionless form and tracking particles in a deformed system of coordinates~\cite{shikhov_perturbation_1960}. An attempt at generalizing this theory to localized geometrical perturbations was made in~\cite{abramov_calculation_1998} but they provide no validation of their technique. It has been recently revisited in \cite{mikolaj_coordinate_change} and~\cite{reed_virtual_2018}, the latter popularizing it under the name of "virtual density theory". It was applied to a fast reactor calculation in a deterministic setting in~\cite{zheng_application_2017}. In particular, it has been shown that the virtual density theory (coupled with regular perturbation theory) and the boundary perturbation theory were equivalent~\cite{reed_virtual_2018_b}. Effectively, for uniform perturbations it allows for the estimation of the change in $k_\text{eff}$ by replacing the geometrical perturbation with a cross-section perturbation. While perturbation theory relying on geometrical changes can be far from straightforward in Monte Carlo, perturbation theory for cross-section perturbations has been extensively studied, and there is a large body of literature investigating the pros and cons of such methods (see~\cite{rief_review_1986, kiedrowski_review_2017} for a review of the field). There are two main classes of Monte Carlo methods for general perturbation theory in reactor physics: adjoint methods, which requires some way of estimating the adjoint flux weighted quantities~\cite{burke_monte_2018,li_calculating_2019}, and the perturbation source method, which explicitly estimates differences in the flux stemming from changes in geometry~\cite{yamamoto_exact_2021,yamamoto_monte_2021}.

Perturbation theory was first applied to virtual density theory in a diffusion, deterministic framework, but was then applied to a transport Monte Carlo simulation for a uniformly perturbed geometry by Yamamoto et al.~\cite{yamamoto_monte_2018}. They used both the differential operator sampling (DOS) method and the correlated sampling method to estimate the first two derivatives of $k_\text{eff}$ with respect to geometrical changes~\cite{rief_generalized_1984,nagaya_estimation_2011,burke_monte_2018}. Non-uniform perturbations (i.e. perturbations modifying space non-uniformly, or only affecting a subspace of the geometry) using virtual density theory in Monte Carlo were first addressed by Dubey et al.~\cite{dubey_uniform_2024}. They propose an algorithm for directly calculating changes in $k_\text{eff}$ due to non-uniform deformations. No perturbation theory was applied, which motivated a recent work from Yamamoto et al.~\cite{yamamoto_monte_2025} where differential operator sampling was applied to non-uniform deformations in a simple multi-group framework. 

In this work, our aim is to investigate one additional class of methods that could supplement these works. The method of intrusive polynomial chaos (gPC), introduced in neutron transport by Poëtte et al.~\cite{poette_gpc-intrusive_2019,poette_efficient_2022}, shows particular promises. In polynomial chaos, the coefficients of a reduced model for e.g. $k_\text{eff}$ are computed, which are essentially a projection of the Monte Carlo solution onto a basis of orthogonal polynomials spanning the space of uncertain parameters. When used in a non-intrusive way, i.e., when the Monte Carlo simulation itself is not modified, numerous calculations are required in order to estimate said coefficients, which is generally undesirable. In~\cite{poette_efficient_2022}, an intrusive algorithm was implemented that uses a time-dependent Monte Carlo solver to emulate a power iteration calculation in order to estimate the coefficients of the reduced model on the fly. In the remainder of this work, we will always refer to intrusive gPC. In~\cite{poette_efficient_2022}, the methodology was mainly used in the context of uncertainty quantification, and, in passing, it demonstrated that it could also be used to represent an uncertain geometrical boundary. While true in theory, using this method to directly consider geometrical perturbations can be daunting in complicated 3D geometries. Therefore, we use the virtual density theory to transform the geometrical perturbation into a cross-section perturbation, in which case the method is straightforward to apply. In stark contrast to perturbation theory, estimating the coefficients of the generalized polynomial chaos model requires that we explicitly simulate the perturbed system. On the other hand, estimating higher order coefficients is straightforward, which in theory may allow estimation of the effect of large perturbations, as well as higher order effects of simultaneous perturbations of geometry and cross-sections.

In this work, we propose a slightly different method for intrusive generalized polynomial chaos (gPC) which is required for static power iteration calculations, and we apply it in a virtual density framework to avoid explicitly perturbing the geometrical parameters. Rather than considering uncertainty propagation, we take advantage of the fact that this method gives, e.g., $k_\text{eff}$ as a function of the system size. We will apply the method to uniform perturbations. This paper is organised as follows: key theoretical background is provided on generalized polynomial chaos and the virtual density theory in Section~\ref{sec:theory}, wherein we also present the details of our method. The accuracy and efficiency of the method is thereafter investigated in Section~\ref{sec:num_results}. Finally, we conclude in Section~\ref{sec:conclusions}.

\section{Theoretical background}
\label{sec:theory}
Before considering the specifics of the necessary Monte Carlo schemes, it is worthwhile to introduce the uncertain counterpart to the linear Boltzmann equation as well as the framework of the generalized Polynomial Chaos. Our method is inspired by the one presented by Poëtte and Brun~\cite{poette_efficient_2022}, although from a different perspective. We also briefly introduce the virtual density theory as theorised by Reed et al.~\cite{reed_virtual_2018}. We only recall what is necessary for a clear presentation of our method.

\subsection{Uncertain Boltzmann equation}
\label{subsec:uncertain_boltzmann}

The uncertain stationary Boltzmann equation driving the motion of neutrons in an uncertain medium is defined for a vector $(\mathbf{r}, \mathbf{\Omega}, E, \mathbf{X})$ in an extended phase space ${\cal P} = \mathbb{R}^6 \times V$, where $V$ is the product space over which the uncertain parameters take their values, i.e. $\mathbf{X}$ is the vector of uncertain parameters. Note that throughout this paper, we use bold font for vectors. The Boltzmann equation for a critical system with angular neutron flux $\phi(\mathbf{r}, \mathbf{\Omega}, E, \mathbf{X})$ reads

\begin{equation}
    \mathbb{L}\phi(\mathbf{r}, \mathbf{\Omega}, E, \mathbf{X}) = \mathbb{F}\phi(\mathbf{r}, \mathbf{\Omega}, E, \mathbf{X}).
    \label{eq:non_norm_stationary}
\end{equation}

We define the net loss operator (the dependency of $\phi$ is left implicit for conciseness)

\begin{equation}
    \mathbb{L} \phi = \mathbf{\Omega} \cdot \nabla \phi + \Sigma_t (\mathbf{r}, E, \mathbf{X}) \phi - \iint \nu_s(\mathbf{r}, E') \Sigma_s(\mathbf{r}, \mathbf{\Omega'} \to \mathbf{\Omega}, E' \to E, \mathbf{X}) \, \phi \, dE' d\mathbf{\Omega'},
\end{equation}

and the fission operator

\begin{equation}
    \mathbb{F} \phi = \frac{\chi(\mathbf{r}, E)}{4 \pi} \iint \nu_f(\mathbf{r}, E') \Sigma_f(\mathbf{r}, E', \mathbf{X}) \,\phi\, dE' d\mathbf{\Omega'}.
\end{equation}

In these definitions, the notations are standard, and fission is assumed isotropic in direction. The only difference with the ``deterministic" Boltzmann equation (i.e. when parameters are deterministic, regardless of how the equation is solved) is the addition of the $\mathbf{X}$ dependency. Note that in this case, we assume that only cross-sections can be uncertain. While slightly less general, we do not need more than that for our purpose, which is to apply this framework to virtual density theory. Eq.\eqref{eq:non_norm_stationary} can be transformed in an eigenvalue problem whose fundamental eigenpair follows

\begin{equation}
    \mathbb{L}\phi(\mathbf{r}, \mathbf{\Omega}, E, \mathbf{X}) = \frac{1}{k_\text{eff}(\mathbf{X})} \mathbb{F}\phi(\mathbf{r}, \mathbf{\Omega}, E, \mathbf{X}).
    \label{eq:eigen_eq}
\end{equation}

$k_\text{eff}(\bm{X})$ is often denoted multiplication factor and gives the expected ratio of neutron production over neutron losses, in this case parametrized by a specific realization of $\bm{X}$. This essentially amounts to normalizing the amount of fission neutrons by the multiplication factor, so that existence of a stationary solution is ensured. 

Eq.\eqref{eq:eigen_eq} can be solved using Monte Carlo power iteration

\begin{equation}
    \mathbb{L}\phi^g(\mathbf{r}, \mathbf{\Omega}, E, \mathbf{X}) = \frac{1}{k_\text{eff}^{g-1}(\mathbf{X})} \mathbb{F}\phi^{g-1}(\mathbf{r}, \mathbf{\Omega}, E, \mathbf{X}),
    \label{eq:pi_eq}
\end{equation}

which amounts to separating neutrons in fission generations, and using the fission neutrons from generation $g-1$ as the source for generation $g$. Then, the fission source (and the flux) will converge to the stationary state as $g$ increases. The generation-wise multiplication factor can be computed using the flux or, equivalently, the fission source:

\begin{equation}
    k_\text{eff}^g(\mathbf{X}) = \frac{\iiint \phi^g(\mathbf{r}, \mathbf{\Omega}, E, \mathbf{X}) d\mathbf{r} d\mathbf{\Omega} dE}{\iiint \phi^{(g-1)}(\mathbf{r}, \mathbf{\Omega}, E, \mathbf{X})d\mathbf{r} d\mathbf{\Omega} dE} = \frac{\iiint n^g(\mathbf{r}, \mathbf{\Omega}, E, \mathbf{X}) d\mathbf{r} d\mathbf{\Omega} dE}{\iiint n^{(g-1)}(\mathbf{r}, \mathbf{\Omega}, E, \mathbf{X})d\mathbf{r} d\mathbf{\Omega} dE},
    \label{eq:keff}
\end{equation}
where $n^{g}(\mathbf{r}, \mathbf{\Omega}, E, \mathbf{X})$ denotes the fission source. Note that in practice, we define $n^{g-1}=n_S^g$ to be the fission source at the start of generation $g$ and $n^g=n_F^g$ as the resulting fission bank at the end of generation $g$, and we recall that if (non-weight preserving) population control is enacted, as is customary for Monte Carlo static power iteration calculations in reactor physics, $n^g$ does not coincide with the fission source used to start the next generation~\cite{mcnp6}. In what follows, we will use the fission source/bank rather than the neutron flux, as it is more practical for our purpose. We will denote $n^g=n_F^g$ the \textbf{fission bank}, i.e. the set of neutrons produced by fission at the end of generation $g$, and $n^{g-1}=n_S^g$ the \textbf{fission source}, i.e. the set of neutrons that serves as the source term for generation $g$. We follow the same convention for all other observables taken at the end (resp. start) of a generation. 

In this context, the usual Monte Carlo machinery appears to hold. In particular, nothing prevents the usual variance reduction and collision biasing techniques from being used.

It is important to remark that in this work, as well as in the works that inspired this investigation~\cite{poette_gpc-intrusive_2019,poette_numerical_2022,poette_efficient_2022,poette_multigroup-like_2023,poette_spectral_2020}, $\mathbb{L}^2$ convergence implies that the eigenpair solution to Eq.~\eqref{eq:eigen_eq} for a given $\mathbf{X}$, converges in some sense to the solution of the deterministic Boltzmann equation for the real system corresponding to this specific realization of $\mathbf{X}$. Therefore, we can use information obtained by solving Eq.~\eqref{eq:eigen_eq} to infer statistical properties of the deterministic Boltzmann equation.

\subsection{Generalized Polynomial Chaos}

Generalized Polynomial Chaos essentially amounts to the decomposition of an arbitrary function on a basis of orthonormal polynomials. It has been successfully used to build reduced models that can be used to accurately but cheaply perform calculations that would otherwise be too expensive. Here, our intent is not to give a rigorous definition of generalized polynomial chaos, but to expose what is necessary for explaining our method. For a more mathematically rigorous perspective, see~\cite{wiener_homogeneous_1938,xiu_wiener--askey_2002,wan_multi-element_2006,blatman_efficient_2010}.

If $\Omega$ is a vector space and $f,g \in \mathbb{L}^2(\Omega)$, we define the scalar product by 

\begin{equation}
    \langle f , g \rangle = \int_\Omega f g ~d\mu,
\end{equation}
with $\mu$ a measure on $\Omega$. We say $(v_k)_{k\in\mathbb{N}}$ is an orthonormal polynomial basis for $\Omega$ with respect to $\mu$ if 

\begin{equation}
    \forall~ p,q \in \mathbb{N}, \quad \langle v_p, v_q \rangle = \delta_{p,q}. 
\end{equation}

Let $f: \Omega \to \mathbb{R}$ a function and $P \in \mathbb{N}$. Then $\forall x \in \Omega$, the gPC approximation of order $P$ of $f$ is given by

\begin{equation}
    f(x) \simeq  \sum_{k=0}^{P} f_k v_k(x) = f^P(x), 
\end{equation}
where the coefficients $(f_k)_{k \in {0,\dots, P}}$ are defined by

\begin{equation}
    f_k = \langle f, v_k \rangle,
\end{equation}
i.e. it is the projection of $f$ on $v_k$. Then $ \forall x \in \Omega, \, \forall f: \Omega \to \mathbb{R}$, $f$ verifies

\begin{equation}
    \sum_{k=0}^{P} f_k v_k(x)  \xrightarrow[P \to +\infty]{{\cal L}^2(\Omega)} f(x),
\end{equation}
where ${\cal L}^2(\Omega)$ denotes the space of square integrable functions. Let us assume we are working in ${\cal P} = \mathbb{R}^6 \times \Omega$, wherein the gPC basis is defined on $\Omega$. In other words, we have $X \in \Omega$, where $X$ is the uncertain parameter, and we want to build a reduced model giving the dependency of $f$ on $X$. Then statistical properties of $f$ can be deduced from the reduced model by defining a functional $I(\phi) : {\cal L}^2({\cal P}) \to {\cal L}^2( \mathbb{R}^6)$ such that

\begin{equation}
    I(\phi) = \int_\Omega R(\phi) d\mu = \mathbb{E}[R(\phi)].
    \label{eq:gpc_functional}
\end{equation}

Depending on the response $R(\phi)$, most usual observables can be computed:
\begin{itemize}
    \item $R(\phi)=\phi$ leads to the flux averaged over possible values of $X$
    \item $R(\phi)=\phi^2$ leads to the 2nd moment of the flux with respect to $X$
    \item $R(\phi)=\phi v_k$ leads to the $k$-th gPC coefficient of the flux
\end{itemize}

Lastly, the choice of the measure $\mu$ and of the polynomial basis depends on the law obeyed by the random parameter. This choice also reflects on the interpretation we can make of the gPC reduced model. In an uncertainty quantification context, we would be interested mostly in the variance due to $X$ or more generally, to sensibility indices for each of the uncertain parameters, in which case the probability law followed by the uncertain parameters should represent the uncertainty on these parameters. This was the approach followed in~\cite{poette_efficient_2022}. In a design context however, we want to use the reduced model to evaluate $f$ for specific values of $X$. It means that {\bf we need to ensure that all values of $X$ are weighted in the same way}. For this reason, we will consider only a uniform law ${\cal U}[\Omega]$, which in 1D is simply ${\cal U}(a,b)$ for $a,b \in \mathbb{R}$. Note that any random variable can be treated as a function of a uniform random variable, hence the implementation of Legendre polynomial is sufficient to consider arbitrary distribution, although in this work we restrict to the case of a uniform variable. When $\mathbf{X}\in\mathbb{R}^{d>1}$, components need to be independent for this property to hold. This is naturally associated with the polynomial basis of Legendre polynomials $P_n$, defined on $[-1,1]$. $f$ must thus be redefined on $[-1,1]$, such that its gPC expansion is given by

\begin{equation}
    f(x) = \sum_{k=0}^{+\infty} f_k P_k\left(\frac{a+b-2x}{a-b}\right), \qquad x \in [a,b],
    \label{eq:gpc_recast_1D}
\end{equation}

and the coefficients are now given by 

\begin{equation}
    f_k = (2 k + 1) \int_{-1}^1 f\left(\frac{b-a}{2}z + \frac{a+b}{2}\right) P_k(z)\,dz.
    \label{eq:gpc_recast_coeffs_1d}
\end{equation}

This can be straightforwardly extended to arbitrary dimension $Q \in \mathbb{N}$, setting $\mathbf{k} = (k_1, \dots, k_Q)$ and $P_\mathbf{k} \to P_{k_1} \times \dots \times P_{k_Q}$, which is a product of 1D Legendre polynomials. Assuming $\Omega = [a_1, b_1] \times \dots \times [a_Q, b_Q]$, Eqs.~\eqref{eq:gpc_recast_1D}-\eqref{eq:gpc_recast_coeffs_1d} then become

\begin{equation}
    f(\mathbf{x}) = \sum_{\mathbf{k}=\mathbf{0}}^{+\infty} f_\mathbf{k} P_\mathbf{k}\left(\frac{a_1+b_1-2x_1}{a_1-b_1},\dots,\frac{a_Q+b_Q-2x_Q}{a_Q-b_Q}\right), \qquad \mathbf{x} \in \Omega,
    \label{eq:gpc_recast_nD}
\end{equation}
and the coefficients are given by 
\begin{equation}
    f_\mathbf{k} = \int_\Omega f\left(\frac{b_1-a_1}{2}z_1 + \frac{a_1+b_1}{2},\dots,\frac{b_Q-a_Q}{2}z_Q + \frac{a_Q+b_Q}{2}\right)  \prod_{i=1}^Q (2 k_i + 1) P_{k_i}(z_i)\,dz_i.
    \label{eq:gpc_recast_coeffs_nd}
\end{equation}

\subsection{Applying generalized polynomial chaos to Monte Carlo power iteration}
\label{sec:combination}
The aim now is to estimate the coefficients of a reduced model for $k_\text{eff}(\mathbf{X})$, where $\mathbf{X}$ is a vector of uncertain parameters. Without loss of generality, we can take a single uncertain parameter, denoted simply by $X\sim{\cal U}[-1,1)$. The Monte Carlo estimate of the fission source (resp. fission bank) $n_t^g(\mathbf{r}, \mathbf{\Omega}, E, X)$ in generation $g$, is given by

\begin{equation}
    n^g_t(\mathbf{r}, \mathbf{\Omega}, E, X) = \sum_{i=1}^{N_{t}^g} \delta(\mathbf{r}) \delta(\mathbf{\Omega}) \delta(E) \delta(X) w_i,
    \label{eq:MCsol}
\end{equation}
where $t\in \{S,F\}$, i.e. a sum of point masses in ${\cal P}$ representing the set of fission neutrons at the end/start of a generation. Using Eq.\eqref{eq:gpc_functional}, we get

\begin{align}
    n_{t}^g(\mathbf{r}, \mathbf{\Omega}, E) &= \int n_{t}^g(\mathbf{r}, \mathbf{\Omega}, E, X) d\mu(X) = \sum_{i=1}^{N_{t}^g} \delta(\mathbf{r_i}) \delta(\mathbf{\Omega}_i) \delta(E_i) \delta(X_i) w_i \label{eq:density_MC}\\
    n_{t}^{k,g}(\mathbf{r}, \mathbf{\Omega}, E) &= \int n_{t}^{g}(\mathbf{r}, \mathbf{\Omega}, E, X) P_k(X) d\mu(X) = \sum_{i=1}^{N_{{t}}^g} \delta(\mathbf{r}_i) \delta(\mathbf{\Omega}_i) \delta(E_i) \delta(X_i) P_k(X_i) w_i,\label{eq:densityGPC_MC}
\end{align}
 and $n_S^g$ (resp. $n_F^g$) denotes the fission source (resp. fission bank) of generation $g$ and its $k$-th gPC coefficients. The fundamental eigenvalue is estimated using the ratio of the population size in the fission bank over the population size in the fission source, which has to be estimated on the fly. Assuming we have converged to the stationary state, we first recall that an estimate of $k_\text{eff}$ stemming from generation $g$ can be written as 

\begin{equation}
    k_\text{eff}^g(X) = \frac{\iiint n_F^g(\mathbf{r}, \mathbf{\Omega}, E, X) d\mathbf{r} d\mathbf{\Omega} dE}{\iiint n_S^g(\mathbf{r}, \mathbf{\Omega}, E, X)d\mathbf{r} d\mathbf{\Omega} dE}
\end{equation}
which has to be projected on the gPC basis, such that

\begin{equation}
    k_\text{eff}^{k,g} = \int k_\text{eff}^g(X) P_k(X) d\mu(X) = \int \frac{\iiint n_F^g(\mathbf{r}, \mathbf{\Omega}, E, X) d\mathbf{r} d\mathbf{\Omega} dE}{\iiint n_S^g(\mathbf{r}, \mathbf{\Omega}, E, X)d\mathbf{r} d\mathbf{\Omega} dE} P_k(X) d\mu(X),
    \label{eq:kgpc}
\end{equation}
and the final Monte Carlo estimates are obtained by averaging those scores over successive generations. 

In the original derivation in~\cite{poette_efficient_2022}, it is stated that Eq.~\eqref{eq:kgpc} needs to be discretised and computed on the fly, which in their case is done using Gaussian quadrature. That is to say, the gPC approximation for $n_F$ and $n_S$ must first be computed and then used to compute the integral over $X$ in Eq.\eqref{eq:kgpc} in order to obtain the gPC approximation for $k_\text{eff}$. For the sake of clarity, we define
\begin{align}
    N^{k,g}_F &= \int d\mu(X)P_k(X)\iiint n_F^g(\mathbf{r}, \mathbf{\Omega}, E, X) d\mathbf{r} d\mathbf{\Omega} dE \nonumber\\
    N^{k,g}_S &= \int d\mu(X)P_k(X)\iiint n_S^g(\mathbf{r}, \mathbf{\Omega}, E, X) d\mathbf{r} d\mathbf{\Omega} dE
    \label{eq:int_est}
\end{align}
the gPC coefficients for the total population size respectively in the fission bank and in the fission source of generation $g$. $N^{P,g}_F$ (resp. $N^{P,g}_F$) then denotes the corresponding gPC model at order $P$. In Eqs.~\eqref{eq:int_est}, the integration is first performed over the usual phase space variables and thereafter integrated against the probability measure for the uncertain variable $X$. Using an intermediate reduced model for $N_F^g$ and $N_S^g$ to compute the polynomial chaos coefficients for $k_\text{eff}$ leads to two remarks: first, performing the outer integration (on $X$) in Eq.\eqref{eq:kgpc} using a Gaussian quadrature introduces an (albeit reasonably negligible) systematic integration error in the result, as a Gaussian quadrature of order $n$ is only exact for polynomials of order smaller than $2n-1$. Second, care needs to be taken to ensure that $N_F^{P,g}$ and $N^{P,g}_S$ are properly converged in polynomial order. Note that for a question of efficiency, provided the stationary state has been reached, $N^{k,g}_F$ and $N^{k,g}_F$ can be replaced by their average up to generation $g$, 

In our case, we made use of the implementation of branchless collisions in Monte Carlo power iteration~\cite{belanger_hunter_effect_2023, bonnet_branchless_2024} to avoid both these issues. Assuming there is no splitting nor Russian roulette, using branchless collision means that each source neutron that does not leak outside the system will produce a single fission neutron with the statistical weight of its parent at the fission event, and the sum of all such neutrons forms the fission source $n_F$. Eq.~\eqref{eq:kgpc} can be simplified by enforcing an unweighted strategy for populating the fission source~\cite{sutton_toward_2022}, ensuring that all neutrons in $n_F$ have unit weight, and applying population control so that the source at the start of a generation, $n_S$, is of constant size $N_S$ across generations, as is commonly done in Monte Carlo power iteration solver for neutron transport~\cite{mcnp6, ROMANO201590,BRUN2015151}. The denominator in the right-hand side of Eq.\eqref{eq:kgpc} then simply becomes $N_S$, the initial population size, and Eq.~\eqref{eq:kgpc} becomes

\begin{equation}
    k_\text{eff}^{k,g} = \frac{1}{N_S}\iiiint n_F^g(\mathbf{r}, \mathbf{\Omega}, E, X) P_k(X) d\mathbf{r} d\mathbf{\Omega} dE d\mu(X).
    \label{eq:kpgc2}
\end{equation}

Plugging Eq.\eqref{eq:densityGPC_MC} into Eq.\eqref{eq:kpgc2}, the gPC coefficients of $k_\text{eff}$ can be directly estimated using 

\begin{equation}
     k_\text{eff}^{k,g} = \frac{1}{N_S} \sum_{i=1}^{N_F^g} \delta(\mathbf{r}_i) \delta(\mathbf{\Omega}_i) \delta(E_i) \delta(X_i) P_k(X_i) w_i,
    \label{eq:keff_est}
\end{equation}
This way, Eq.~\ref{eq:kgpc} is computed without estimating an intermediate reduced model for the fission source, and the statistical error we get on the coefficients of the reduced model for $k_{eff}$ are reliable (as much as statistical errors obtained from Monte Carlo power iteration calculations are reliable~\cite{brissenden_biases_1986}). Note that the use of Eq.~\eqref{eq:keff_est} requires that all neutrons in $n_S^g$ start with unit weight and that $N_S$ is constant across generations. In the rest of this work, we used both the estimator given by Eq.\eqref{eq:keff_est} and the estimator introduced in~\cite{poette_efficient_2022}, and we checked that they gave consistent results for the average value of the polynomial chaos coefficients. For the sake of clarity, in figures we only represent one of the two estimators. We have observed that the estimator introduced in~\cite{poette_efficient_2022} tends to yield lower statistical uncertainties on the estimates of the gPC coefficients for $k_\text{eff}$. In the following, $k_\text{eff,1}$ (resp. $k_\text{eff,2}$) denotes estimates obtained using Eq.\eqref{eq:keff_est} (resp. the estimator introduced in~\cite{poette_efficient_2022}). 

Lastly, we want to stress a difference between our algorithm and the one developed in~\cite{poette_efficient_2022} for dynamic power iteration. In the latter, the authors solve the unstationary Boltzmann equation for the time-dependent flux, or equivalently, the time-dependent neutron density defined by $n(\mathbf{r}, \mathbf{\Omega}, t, X)$. The solution is evaluated at specific time-steps $t^n=t^{n-1}+\Delta t,~n\in\{0,\dots,T\}~\text{and}~T\in\mathbb{N}$ with a fixed time-step size $\Delta t\in\mathbb{R}$. These time-steps are formally defined as "generations", but they are not \textbf{fission generations}: that is to say $n(\mathbf{r}, \mathbf{\Omega}, t^n, X)$ is simply the distribution of particles in the extended phase-space  ${\cal P}$ at time $t^n$, which differs from the distribution of fission particles at time $t^n$, the latter not even being clearly defined in a time-dependent context. Asymptotically thought, we get

\begin{equation}
    k_\text{eff,dyn}^n=\frac{\iiint n(\mathbf{r}, \mathbf{\Omega}, t^n,X) d\mathbf{r}d\mathbf{\Omega}dE}{\iiint n(\mathbf{r}, \mathbf{\Omega}, t^{n-1},X) d\mathbf{r}d\mathbf{\Omega}dE}.
\end{equation}
It is well known that $k_\text{eff, dyn}$ obtained through dynamic power iteration, and its associated eigenfunction, only coincide with $k_\text{eff}$ and its eigenfunction around criticality, where $k_\text{eff}$ is obtained through static power iteration as defined in Eq.~\eqref{eq:keff}~\cite{cullen_static_2003}. Consequently, gPC models for $k_\text{eff, dyn}$ and $k_\text{eff}$ can only coincide near criticality. In addition, the gPC model for $n(\mathbf{r}, \mathbf{\Omega}, t^n,X)$ in~\cite{poette_efficient_2022} is time-dependent, i.e. there are as many models as there are time-steps/generations. At each time-step, the gPC models are computed by integrating against the probability measure $\mu_{t=0}$ from which the uncertain parameter $X$ was initially sampled from, that is to say, there is a single source term, which is given by the initial condition of the time-dependent calculation.

\begin{figure}
    \centering
    \includegraphics[width=0.5\linewidth]{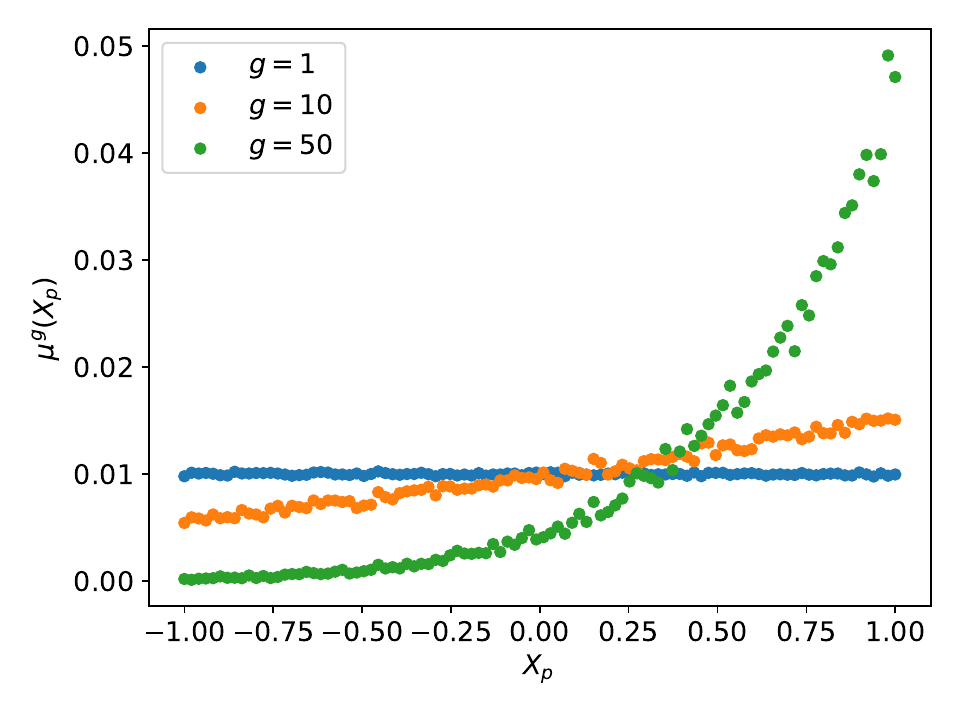}
    \caption{Probability distribution of the uncertain parameters in $n_S^g$ for the box system after different numbers of generations $G$ when the Duplicate-Discard method is used.}
    \label{fig:non_stationary_dist}
\end{figure}

In the static power iteration that our algorithm is based on, each generation can be considered as an independent simulation with a source term that is sampled from the fission bank of the previous generation\footnote{As explained in~\cite{brissenden_biases_1986}, this is obviously not true, as samples obtained from successive generations are correlated, but Monte Carlo scores in power iteration codes are generally accrued under this hypothesis.}. In a regular static power iteration, a stationary state will be reached after generation $G_d$, such that $\forall g>G_d$
\begin{align}
    n_S^g(\mathbf{r}, \mathbf{\Omega},E)\simeq n_S^{g-1}(\mathbf{r}, \mathbf{\Omega},E)\\
    n_F^g(\mathbf{r}, \mathbf{\Omega},E)\simeq n_F^{g-1}(\mathbf{r}, \mathbf{\Omega},E),
\end{align}
and each subsequent generation yields a sample of $n_S^g, n_F^g$, and $k_\text{eff}^g$. Usually, the generation dependency is suppressed, using the stationarity assumption, and all generations (after convergence) contribute to the same tally, in contrast with an unstationary simulation for which tallying needs to remain time-dependent. Now let us turn to the gPC calculation based on the static power iteration algorithm. We define $\mu^g(X)=\mathbb{P}(X_i=X| n_i\in n_S^g)$, where $n^g_S$ denotes the fission source for generation $g$ and $n_i$ is a neutron with phase-space coordinates $(\mathbf{r}_i,\mathbf{\Omega}_i,E_i,X_i)$. Said otherwise, $\mu^g(X)$ is the probability that, selecting a particle $n_i$ in the fission source for generation $g$, it has uncertain parameter $X_i = X$. Fig.~\ref{fig:non_stationary_dist} shows that, for the uniform expansion of the box system that will be investigated in Section~\ref{sec:num_results}, when the Duplicate-Discard population control method is used~\cite{leppanen_serpent_2013}, the probability distribution $\mu^g$ is not stationary. We stress here that $\mu^{g=0}=\mu$, where $\mu$ is the (uniform) probability distribution from which the uncertain parameters in the first generation are initialized; it is the distribution of uncertain parameters conditioned to being in the fission source of generation $g$ that is non-stationary. This is explained by the fact that in this case, $k_\text{eff}(X)$ is a decreasing function of $X,~\forall X \in [-1,1]$. Therefore, less fission progenies carrying small values of $X$ are added to $n_F^g$, which leads to smaller values of $X$ being under-represented in the fission source $n_S^{g+1}$ for the next generation. Because $n_S^{g+1}$ acts as the source term for the next generation, strictly speaking this means that the integration over $\mu$ in Eqs.~\eqref{eq:density_MC},\eqref{eq:densityGPC_MC},\eqref{eq:kgpc}, and \eqref{eq:int_est} needs to be replaced by an integration over $\mu^g$. Similarly, the polynomial basis needs to become generation-dependent and determined on the fly, as it needs to be orthogonal with respect to $\mu^g$, which is unknown a priori. Consequently, a different gPC model needs to be computed for each generation, which, in substance, would make the algorithm very similar to~\cite{poette_efficient_2022}, with the difference that in~\cite{poette_efficient_2022}, once convergence of $k_\text{eff,dyn}$ is achieved, the associated polynomial model (for $k_\text{eff,dyn}$, not the population density) can be averaged over successive time-steps, as the polynomial basis remains the same throughout the calculation. This would not be possible in our case, as nothing ensures that the polynomial basis associated with $\mu^g$, and therefore the gPC coefficients, will reach a stationary state. Moreover, existence of a marginal stationary state, i.e. stationarity in the usual phase-space coordinates, but not in the uncertain dimension, is unclear and would need to be proven. It can also be expected to be detrimental in terms of statistical convergence: provided the power iteration calculation is started with $N$ particles, over $G_a$ active generations (i.e. scoring generations after an eventual marginal stationary state has been reached), we would have $G_a$ models each estimated with $N$ particles, instead of a single model estimated with $N\times G_a$ particles. For these reasons, we have chosen to leave this venue to future work, and we follow an alternative approach.

Indeed, one may address the non-stationarity of $\mu^g$ by sampling the particles in $n_S^{g+1}$ from those in $n_F^{g}$ in such a way that $\mu^g$ is forced to be stationary and uniform, i.e. $\forall ~g, \mu^g =\mu$. To do so, let us define $\nu^g(X)=\mathbb{P}(X_i=X|n_i\in n_F^g)$. It is the probability that, selecting particle $n_i$ in the fission bank at the end of generation $g$, it has uncertain parameter $X_i=X$. We stress that we use $\nu^g$ for the probability distribution of $X$ in $n_F^g$ and we use $\mu^g$ for the probability distribution of $X$ in $n_S^g$. We use 
\begin{equation}
    \psi(X) = 1/\nu^g(X)
\end{equation}
as the importance function in a variant of the importance combing procedure given in~\cite{thomas_e_booth_weight_1996}. The algorithm is written in a mathematically rigorous way in Appendix~\ref{appendix1}, and summarized in Algorithm~\ref{algo:alg1}. The combing is applied to select $N_S$ neutrons in $n_F^g$, which are added to $n_S^{g+1}$ and form the fission source for generation $g+1$. Note that this modification makes this procedure non weight-preserving, and that we assume that all particles before combing have unit weight (which is ensured in our algorithm). Its output is a population $n_S^{g+1}$ of $N_S$ particles sampled from $n_F^g$ with unit weight, such that $\mu^{g+1} =\mu^g=\mu$. Note that the possibility of population control algorithms targeted at the uncertain parameter is already hinted at in~\cite{poette_efficient_2022}. However, in their algorithm, the population control needs to be unbiased at least for the first statistical moment, on the entire extended phase space ${\cal P}$. Said otherwise, population control in that case is only seen as a possible variance reduction method, as is customary in time-dependent simulations. The initial importance combing introduced in~\cite{thomas_e_booth_weight_1996} for instance could be used for that purpose, although these investigations are outside the scope of this work.

The distribution $\nu^g$ is unknown a priori and needs to be approximated on the fly. This can be done in numerous ways. When $dim~\mathbf{X} = 1$, we used a polynomial fit whose coefficients are obtained by using a least square fitting procedure on the histogram of the $X$s in $n_F^g$. In that case the least square fitting was performed using the gels procedure of the LAPACK library~\cite{anderson1999lapack}. When $dim~\mathbf{X} > 1$, we chose to directly use the histogram of the $X$s of the particles in $n_F^g$. In addition, we have observed that the order of the polynomial fit needs to be at least as high as that of the gPC decomposition for the latter to give accurate results. While this phenomenon has not been investigated in details, we think this is explained by the fact that fitting a probability distribution using a polynomial fitting procedure of order $m$ by the least square method only preserve the statistical moments of the probability distribution up to order $m$~\cite{method_samejima_1979}. One should also ensure that the initial population size is large enough to ensure that the space of uncertain parameters is adequately covered in order to avoid introducing an undersampling bias.

Note that in this combing procedure, all particles sharing the same value of $X$ have an equal chance of being selected, i.e., for fixed $X$ the procedure should not bias the distribution of usual phase space coordinates, although it modifies the distribution of particles in the extended phase-space ${\cal P}$ (which includes the uncertain space), by definition. While our numerical results do not appear to suffer from a bias when compared to reference calculations (as will be shown in Section~\ref{sec:num_results}), more detailed analysis of this step would be welcomed to improve the reliability of the method, and will be the subject of future works. Alternatively, the use of more robust moments preserving techniques should be investigated to approximate $\nu^g$, and is likewise left for future works. 

\begin{algorithm}
    \caption{Importance combing for enforcing that $\mu(X)$ be a uniform law}
    \begin{algorithmic}[1]
        \Procedure{importanceCombing}{$\nu, \mathbf{n}_S, \mathbf{n}_F, N_S$}
        \State Shuffle $\mathbf{n}_F$ to avoid bias
        \State Reinitialize $\mathbf{n}_S$
        \State $U \gets 0$ 
        \Comment{Initialize total importance}
        \For{$p'_i$ in $\mathbf{n}_F$}
            \State $I_i \gets 1 / \nu(X'_i)$
            \Comment{Compute the importance}
            \State $U {+}{=} I_i$
            \Comment{Compute the total importance}
        \EndFor
        \State $r \sim {\cal U}[0,1)$
        \State $\xi = r \frac{U}{N}$ 
        \Comment{Offset to avoid bias}
        \State $C = 0$
        \For{$p'_i$ in $\mathbf{n}_F$}
            \State $C~ {+}{=}~ I_i$
            \While{$\xi < C$}
                \State $\mathbf{n}_S \gets p'_i$
                \Comment{particle $p'_i$ is copied and added to $n_S$}
                \State $w'_i = 1$
                \Comment{weight of the selected particle is set to one}
                \State $\xi ~{+}{=}~ U/N_S$
            \EndWhile
        \EndFor
        \State \Return $\mathbf{n}_S$
        \Comment{Output the source for next generation}
        \EndProcedure
    \end{algorithmic}
    \label{algo:alg1}
\end{algorithm}

Our Monte Carlo algorithm for the resolution of the uncertain Boltzmann equation in stationary problems is provided in Algorithm~\ref{algo:alg2}. $G_d$ is the number of inactive generations necessary for the system to converge to the stationary state starting from an arbitrary initial condition. $G_a$ is the number of active generations over which the tallies are estimated, namely the gPC coefficients for $k_{\text{eff}}$ and $N_F$. $P$ is the polynomial order for the gPC expansion, and $N_S$ is the initial population size, assuming all neutrons start with unit weight. In the following, we explicate each significant step of Algorithm~\ref{algo:alg2}:

\begin{itemize}
    \item In step 2, we initialize an arbitrary initial guess for the fission source. Particles are assigned with an uncertain parameter $X$ sampled from the uniform law ${\cal U}[-1,1)$.
    \item In step 3, we initialize a placeholder for the fission neutrons produced in a generation
    \item In steps 7 and 9, we use usual algorithms for sampling flight distance and collision channel. Note that in each of those steps, sampling procedures may depend explicitly on $X$, which essentially plays the role of a regular phase-space variable. This is in stark contrast with perturbation theory, in which the effect of perturbations is often computed only through tallies, and not through explicit modifications of the stochastic process. Fission neutrons are placed in $\mathbf{n}_F$. Note that fission neutrons inherit the value of $X$ from their parent neutron, whereas regular coordinates are sampled in the usual way. 
    \item If the current generation is later than the discard criterion $G_d$, in step 15, we tally Eq.~\eqref{eq:int_est}. In step 16, we tally Eq.~\eqref{eq:keff_est}, and finally, in step 19, we integrate Eq.~\eqref{eq:kgpc} using Gaussian quadrature of order $n\geq P$ and accumulate the resulting coefficients for $k_{\text{eff},2}$. $X_n$ is the abscissa of the $n$-th Gaussian quadrature point and $W_n$ its weight. For each observable, contributions from all generations later than $G_d$ are accumulated in the same tally.
    \item In step 22, we approximate the distribution $\nu^g$ of the random parameter $X$ in the fission neutrons held in $\mathbf{n}_F^g$. As mentioned above, when $dim~\mathbf{X} = 1$, we use a polynomial fitting procedure. Otherwise, we approximate $\nu^g$ by its histogram approximation. 
    \item In step 23, $\mathbf{n}_S^{g+1}$ for the next generation is sampled using Algorithm~\ref{algo:alg1}.
\end{itemize}

\begin{algorithm}
    \caption{Main loop for the Monte Carlo implementation}
    \begin{algorithmic}[1]
        \Procedure{run}{$N, P, G_d, G_a$}
        \State Initialize starting fission source $\mathbf{n}^0_S=\{p_1, \dots, p_{N_S}\}$
        \Comment{$p_i$'s have fields $\{\mathbf{r}_i,\mathbf{\Omega}_i,E_i,X_i,w_i\}$}
        \State Initialize holder for final fission source $\mathbf{n}_F^g$
        \Comment{$p'_i$'s in $\mathbf{n}_F$ have fields $\{\mathbf{r}'_i,\mathbf{\Omega}'_i,E'_i,X'_i,w'_i\}$}
        \While{$g < G_a + G_d$}
            \For{$p_i$ in $\mathbf{n}_S^g$}
                \While{$p_i$ is alive}
                    \State doTransport$(p_i)$
                    \If{collide} 
                        \State doCollision$(p_i,\mathbf{n}_F^g)$
                    \EndIf
                \EndWhile
            \EndFor
            \If{$g > G_d$}
                \Comment{We accumulate tallies once converged to the stationary state}
                \For{k=1,P}
                    \State $N_F^{k} {+}{=} \sum_{p'_i \in \mathbf{n}_F} w'_i P_k(X'_i)$
                    \State $k_\text{eff,1}^{k} {+}{=} \left(\sum_{p'_i \in \mathbf{n}_F} w'_i P_k(X'_i)\right) / {N_S}$
                \EndFor
                \For{k=1,P}
                    \State $k_\text{eff,2}^{k} {+}{=} \left(\sum_n N_F^{P}(X_n) W_n\right)/{N_S}$
                \EndFor
            \EndIf
            \State approximateProbDistribution$(\nu^g, \mathbf{n}_F^g)$   
            \State $\mathbf{n}_S = $ importanceCombing$(\nu^g, \mathbf{n}_F^g, \mathbf{n}_S^{g+1}, N_S)$
            \State Clear $\mathbf{n}_F^g$
            \EndWhile
        \EndProcedure
    \end{algorithmic}
    
    \label{algo:alg2}
        
\end{algorithm}

In this framework, it should be clear that customary variance reduction techniques should remain valid. This can be seen by rewriting Eq.~\eqref{eq:eigen_eq} in its integral form and applying the formalism developed in~\cite{lux_monte_1991}. We do not provide such a derivation here, as it is straightforward. Interestingly, this also opens the door for variance reduction techniques targeting the uncertain parameters. Although this has not been investigated here, as our distribution of uncertain parameters is simple, it may be of interest when considering experimental uncertainties on cross-sections for which the probability distribution is non-trivial and may exhibit long tails.

Finally, it has been shown that generalized polynomial chaos for the Boltzmann equation with uncertain parameters converges quickly with polynomial order~\cite{poette_spectral_2020}, and its speed of convergence has been compared with that of perturbation theory in~\cite{poette_efficient_2022}, and shown to be faster for some simple problems. Additionally, it is worth making two interesting remarks:

\begin{itemize}
    \item While higher order perturbation theory in Monte Carlo requires non-trivial calculations and implementation, increasing the gPC order only requires a higher order Legendre polynomial, which is easy to compute. Additionally, the theory for extending internal boundary perturbation theory to higher orders is not so clear.
    \item Contrary to perturbation theory, gPC models are subjected to the curse of dimensionality, i.e. the number of coefficients to score increases exponentially with the number of uncertain parameters, as the number of coefficients to compute is $(P+1)^Q$. Note that in our case, as the dimension of the uncertain space remains small, the curse of dimensionality will not be much of an issue. In fact, the computational cost will be comparable to that of a regular estimator for $k_\text{eff}$. 
\end{itemize}

\subsection{The virtual density theory with generalized polynomial chaos}
\label{virtual_density_theory}

While treating geometrical perturbations with generalized polynomial chaos is possible in theory, it is impractical in practice: it requires tracking all the perturbed surfaces for each particle, the crossing of which becomes hard to follow in complex geometries. It is possible to counteract this limitation by using the framework of virtual density. In a nutshell, the virtual density framework consists in equating geometrical perturbations with cross-section perturbations. This principle is almost as old as the Monte Carlo method~\cite{shikhov_perturbation_1960}, but was revisited recently by Reed et al.~\cite{reed_virtual_2018} and Kowalski et al.~\cite{mikolaj_coordinate_change}. For more details on the virtual density framework we refer the reader to their work; here we only recall what is necessary for explaining the proposed method, and we present our Monte Carlo algorithm. 

\subsubsection{Reminder on the virtual density theory}
The virtual density framework is able to model uniform and non-uniform, isotropic or anisotropic geometrical deformations, each of those terms having a specific meaning in the context of virtual density:
\begin{itemize}
    \item Uniform perturbations refer to geometrical perturbations of constant magnitude throughout the geometry. Conversely, a non-uniform perturbation refers to one that is either localised or whose amplitude varies within the geometry.
    \item Isotropic perturbations are invariant by rotation. Conversely, anisotropic perturbations depend on the direction.
    \item An expansion is a geometrical perturbation without mass conservation. A swelling refers to a geometrical perturbation with mass conservation.
    \item Negative expansion or swelling refers to a situation where linear lengths are decreased, while positive expansion or swelling refers to a situation where linear lengths are increased.
\end{itemize}

Applying virtual density essentially amounts to stretching the neutron path when it evolves in a perturbed region. In the general case, the neutron path has to be stretched anisotropically, which can be done straightforwardly by adequately modifying the outgoing flight direction/distance after colliding in a perturbed region. This is the only change required to a Monte Carlo code if the perturbation is uniform~\cite{yamamoto_monte_2018}. On the other hand, explicitely simulating non-uniform perturbation requires one to go back to the framework of coordinate changes initially introduced in ~\cite{shikhov_perturbation_1960} and recently revisited in~\cite{mikolaj_coordinate_change}, and requires that more significant changes be made to the tracking algorithm, and those will not be considered in this paper but left for future works. 

Let us stress here that in our case, we are directly simulating the modified flight path, which means that neutrons are effectively evolving in the modified geometry, and there is \textbf{no} perturbation theory applied in this case. This is in stark contrast with the algorithm developed in~\cite{yamamoto_monte_2018} in which neutrons evolve in the unperturbed geometry and the effect of the geometrical change is estimated using perturbation tallies.

In order to fix the notations, let $V$ be a box with linear length $\mathbf{L} = (L_x, L_y, L_z)$, and $V'$ the perturbed box with linear length $\mathbf{L'} = (f_x L_x, f_y L_y, f_z L_z)$. We denote $\mathbf{f} = (f_x, f_y, f_z)$ the deformation factor of the geometrical perturbation. A neutron undergoes a flight $\mathbf{l} = (l_x, l_y, l_z)$ with unit direction vector $\mathbf{u} = \mathbf{l}/||\mathbf{l}||$, and we denote $\mathbf{l'}$ the perturbed flight with its perturbed direction vector $\mathbf{u'}$. $||\mathbf{l}||$ is sampled in the usual way, so that
\begin{equation}
    ||\mathbf{l}||=-\frac{1}{\Sigma_t(\mathbf{r}, \mathbf\Omega,E)} ln(1-u),
\end{equation}
where $u \sim{\cal U}[0,1)$ and $\Sigma_t(\mathbf{r}, \mathbf\Omega,E)$ is the total macroscopic cross-section. A general Monte Carlo algorithm using surface tracking can be implemented in a few steps: 
\begin{enumerate}[label=(\roman*)]
    \item $\mathbf{l'}$ is computed depending on the perturbation kind (swelling or expansion) and $\mathbf{f}$.
    \item The flight stretching factor $s = ||\mathbf{l'}|| / ||\mathbf{l}||$ is computed. 
    \item $\mathbf{u'}$ is computed using $s$ and $\mathbf{f}$, and the particle is moved by $||\mathbf{l'}||$ along the perturbed direction. 
    \item Before collision or surface crossing, the particle recovers its unperturbed direction, in order to preserve angular distributions
\end{enumerate}

When the perturbation is an expansion, $\mathbf{l'}$ and $\mathbf{u'}$ are respectively given by

\begin{equation}
    \mathbf{l'} = \left(\frac{l_x}{f_x}, \frac{l_y}{f_y}, \frac{l_z}{f_z}\right)~\text{and}~\mathbf{u'} = \left(\frac{u_x}{s f_x}, \frac{u_y}{s f_y}, \frac{u_z}{s f_z} \right).
    \label{eq:f_expansion}
\end{equation}

When the perturbation is a swelling, they are respectively given by

\begin{equation}
    \mathbf{l'} = \left(l_x f_y f_z, l_y f_x f_z, l_z f_x f_y\right)~\text{and}~\mathbf{u'} = \left(\frac{u_x f_y f_z}{s}, \frac{u_y f_x f_z}{s}, \frac{u_z f_x f_y}{s} \right).
    \label{eq:f_swelling}
\end{equation}

\subsubsection{Generalized polynomial chaos applied to virtual density}

The modifications needed to combine virtual density with generalized polynomial chaos are minimal. Step 7 in Algorithm~\ref{algo:alg2} needs to be modified to account for the fact that the sampling of neutron flights depends on $\mathbf{X}$. Regarding the procedure described in the previous subsection, $\mathbf{X}$ is related to $\mathbf{f}$ so that

\begin{equation}
    f_i = \frac{L'_i}{L_i} = 1 + \epsilon_i X_i,
\end{equation}
where $i \in \{x,y,z\}$ and $X_i \sim {\cal U}[-1,1]$, and each neutron will see its path deformed differently. 

We need to clarify that choosing $\epsilon_x = \epsilon_y = \epsilon_z$ does not mean that we are necessarily in the case of isotropic virtual density. $\mathbf{\epsilon}$ indicates how much of the space of stretching parameters we explore, and the isotropic case is achieved by sampling a single value for all directions, while the anisotropic case is achieved by sampling different values for different directions. As the efficiency of generalized polynomial chaos depends on the size of the uncertain space to be explored, $\mathbf{\epsilon}$ should be chosen so that it avoids sampling regions in the uncertain space that are not relevant. In our case, we shall be interested in isotropic ($\epsilon_x X_x=\epsilon_y X_y= \epsilon_z X_z$), radial ($\epsilon_x X_x=\epsilon_y X_y\neq0$ and $\epsilon_z X_z =0$) and axial ($\epsilon_x X_x=\epsilon_y X_y=0$ and $\epsilon_z X_z \neq0$) perturbations.

$\mathbf{f(\mathbf{X})}$ is now particle specific, and the virtual density algorithm presented in the previous section can be straightforwardly used with no changes, aside from using particle-wise stretching factors instead of global stretching factors. In particular, Eqs.\eqref{eq:f_expansion}-\eqref{eq:f_swelling} can be adapted simply by replacing all occurrence of $f_i$ by $f_i(X_i)$.

\section{Numerical Results}
\label{sec:num_results}

We first attempt to demonstrate the method by applying it to a simple fuel box, with nuclide densities given in Table~\ref{tab:box_densities}. The (unperturbed) fuel box has width $80~\text{cm}\times80~\text{cm}\times80~\text{cm}$. Leakage boundary conditions are applied on all sides. This system is not intended to be realistic, but simply to estimate the accuracy of the method in a simple benchmark. The next problem is a reflector problem with the previous box immersed in a $100~\text{cm}\times100~\text{cm}\times100~\text{cm}$ box filled with water, in order to verify that the method is able to address properly indirect effects. Leakage boundary conditions are applied on all sides. Finally, we turn to the more realistic case of a MOX fuel assembly of the C5G7 benchmark~\cite{nea_benchmark_2000}. In this reference, the fuel assembly is in 2D, but we chose to extend it to 3D with a height of $100~\text{cm}$ in the $z$ direction. Periodic boundary conditions are applied in the $x$ and $y$ directions, and leakage boundary conditions are applied in the $z$ direction. The  multiplication factor for each system when no perturbation is applied is given in Table~\ref{tab:keff}.

\begin{table}[]
    \centering
    \begin{tabular}{|c|c|c|}
    \hline
    Isotope & Fuel & Water \\\hline
    $^{1}\textbf{H}$ & - & \num{9.00636e-2}  \\
    $^{16}\textbf{O}$ & \num{4.50318e-2} & \num{4.50318e-2}  \\
    $^{235}\textbf{U}$ & \num{5.72863e-4} & -  \\
    $^{238}\textbf{U}$ & \num{1.28292e-2} & - \\
    $^{238}\textbf{Pu}$ & \num{1.46306e-4} & - \\
    $^{239}\textbf{Pu}$ & \num{7.23300e-4} & -\\
    $^{240}\textbf{Pu}$ & \num{1.20500e-3} & -\\
    $^{241}\textbf{Pu}$ & \num{3.33022e-4} & -\\
    $^{242}\textbf{Pu}$ & \num{4.18070e-4} & -\\
    $^{241}\textbf{Am}$ & \num{3.15180e-5} & -\\\hline
\end{tabular}
    \caption{Isotopic compositions for the box systems (in $\text{b}^{-1}~\text{cm}^{-1}$).}
    \label{tab:box_densities} 
\end{table}

\begin{table}[]
    \centering
    \begin{tabular}{c|c}
        System & $k_\text{ref}$ \\ \hline
        Box & $0.96017\pm6$~pcm\\
        Reflected box & $1.06617 \pm 10$~pcm \\ 
        Assembly & $1.14069 \pm 9$~pcm
    \end{tabular}
    \caption{Reference multiplication factor for each of the investigated systems.}
    \label{tab:keff}
\end{table}

\subsection{Numerical results for the box systems}

\begin{figure}
    \centering
    \begin{subfigure}{0.45\textwidth}
        \includegraphics[width=\textwidth]{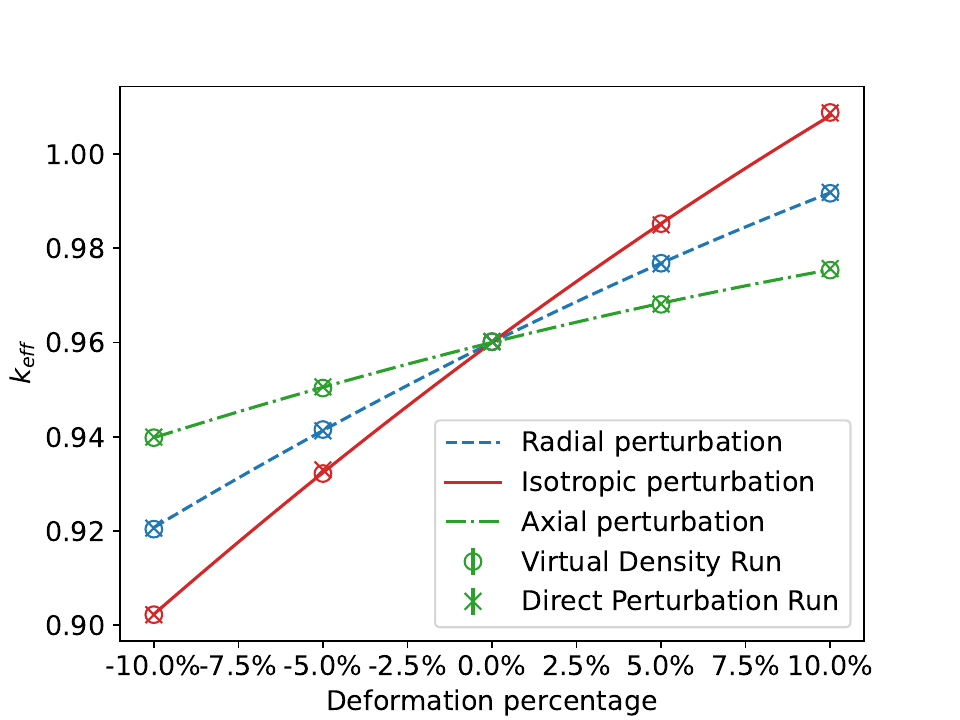}
        \caption{Uniform expansion}
        \label{fig:uni_box_exp}
    \end{subfigure}
    \begin{subfigure}{0.45\textwidth}
        \includegraphics[width=\textwidth]{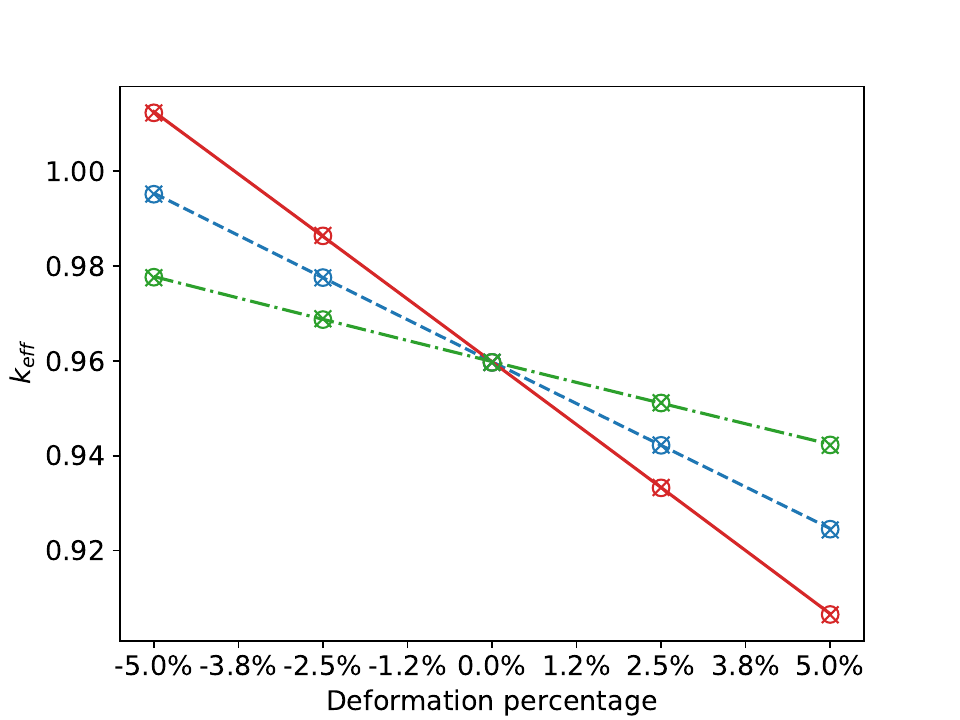}
        \caption{Uniform swelling}
        \label{fig:uni_box_swl}
    \end{subfigure}
    \caption{Numerical comparison between virtual density, direct calculations, and gPC using virtual density for uniform swelling or expansion of the box system. Full lines represent the (continuous) gPC model, circles stand for virtual density and crosses for direct perturbations.}
    \label{fig:uni_box}
\end{figure}

Results for the box system are shown in Fig~\ref{fig:uni_box}. $2$nd order in Legendre polynomials was sufficient for gPC results to agree with reference runs within statistical uncertainty. $G_d=100$ inactive generations were sufficient to reach stationary state, and the gPC coefficients were obtained with $G_a =10^3$ active generations and $N_S=10^6$ neutrons. Note that because the coefficients of the gPC model are obtained through Monte Carlo, they also come with statistical uncertainties. Additionally, due to the low dimension of our uncertain space, scoring gPC coefficients adds little computational time. Note that in the following cases, we performed one gPC calculation for each kind of perturbation (i.e. a single calculation for all possible isotropic expansions, another for all possible radial expansions, etc.), so that the gPC model is always single-dimensional. This is not a requirement of the method, but as we were not interested in simultaneous radial and axial perturbations, it was more computationally efficient to proceed in this way. More precisely, it means that each line in the following figures corresponds to an independent gPC calculation. 

In Figure \ref{fig:uni_box}, the values of $k_\text{eff}$ predicted by the gPC models are compared against reference simulations realized by directly altering the geometry (described as direct perturbation runs) and by using the virtual density theory - but no perturbation theory (described as virtual density runs). All virtual/direct Monte Carlo results were obtained with $1\sigma$ uncertainty below $10~\text{pcm}$, which is almost indiscernible on these plots. All calculations results are within statistical uncertainty of one another. For swelling, the sampling range for deformations has been chosen to be smaller than for expansion, in order to keep reactivity changes roughly of the same magnitude. In this case, given that the system is symmetric in all $3$ directions, the effect of geometrical perturbations depends not on the direction of the perturbation, but on the number of perturbed directions. 

We define the reactivity changes by:
\begin{align}
    \Delta\rho_\text{i} &= 1/k_\text{ref} - 1/k_\text{i}
    \label{eq:deltaRho}
\end{align}
where $k_\text{ref}$ is the multiplication factor when no perturbation is applied (see Table~\ref{tab:keff}), and $i$ stands for `virtual'. `real' or `gPC' results. $k_i$ is the multiplication factor obtained respectively through the virtual density reference runs, the direct perturbation runs, and predicted by the corresponding gPC model, for a given perturbation. $\delta(\Delta\rho_\text{virtual})$ (resp. $ \delta(\Delta \rho_\text{gpc})$) is the relative error on $\Delta \rho_\text{virtual}$ (resp. $\Delta\rho_\text{gpc}$) compared to $\Delta\rho_\text{real}$. It is defined by
\begin{equation}
    \delta(\Delta \rho_\text{virtual}) = \frac{\Delta\rho_\text{virtual}-\Delta\rho_\text{real}}{\Delta\rho_\text{real}},
\end{equation}
and similarly for $\delta(\Delta \rho_\text{gpc})$, the error on the reactivity change predicted by the gPC model relative to the reactivity change computed by direct perturbations. Some values of $\Delta\rho$ and its errors are reported in Table~\ref{tab:err_uni_exp_box} for the isotropic expansion. These details are provided only for isotropic deformations for the sake of conciseness, as radial and axial perturbations follow the same trend, as evidenced by Fig.~\ref{fig:uni_box}. The relative errors between virtual density or gPC predictions and direct perturbation results are under $1\%$ and generally lie within statistical uncertainty, even for large geometrical perturbations. 

\begin{table}[]
    \centering
    \begin{tabular}{ c c c c c c c }
    \hline
    {\bf Perturbation} & $\Delta\rho_\text{direct}$ & $\Delta\rho_\text{virtual}$ & $\delta(\Delta\rho_\text{virtual})$ & $\Delta\rho_\text{gPC}$ & $\delta(\Delta\rho_\text{gPC})$ \\ \hline
    $-10\%$ & $-6691 \pm 12$ & $-6688 \pm 12$ & $-0.04\%$ & $-6704$ & $0.2\%$ \\ 
    $-5\%$ & $-3124 \pm 12$ & $-3126 \pm 12$ & $0.07\%$ & $-3123$ & $-0.02\%$ \\ 
    $0\%$ & 0 & 0 & 0 & $-42$ & $-0.05\%$ \\ 
    $5\%$ & $2673 \pm 12$ & $2659 \pm 12$ & $0.12\%$ & $2676$ & $-0.16\%$ \\ 
    $10\%$ & $5015\pm 12$ & $5024 \pm 12$ & $-0.5\%$ & $5007$ & $0.19\%$ \\ 

    \hline
\end{tabular}
    \caption{Relative errors on reactivity changes due to uniform expansion of the box system. Results are given in $\text{pcm}$}
    \label{tab:err_uni_exp_box} 
\end{table}

\begin{figure}
    \centering
    \begin{subfigure}{0.45\textwidth}
        \includegraphics[width=\textwidth]{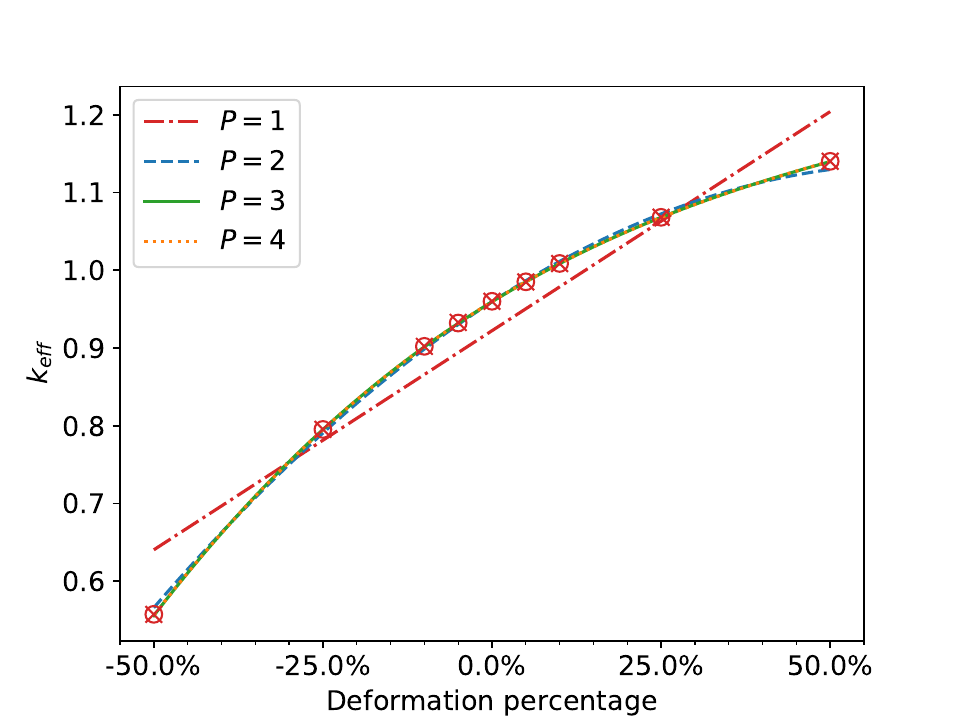}
        \caption{Convergence in polynomial order.}
        \label{fig:uni_box_conv_P}
    \end{subfigure}
    \begin{subfigure}{0.45\textwidth}
        \includegraphics[width=\textwidth]{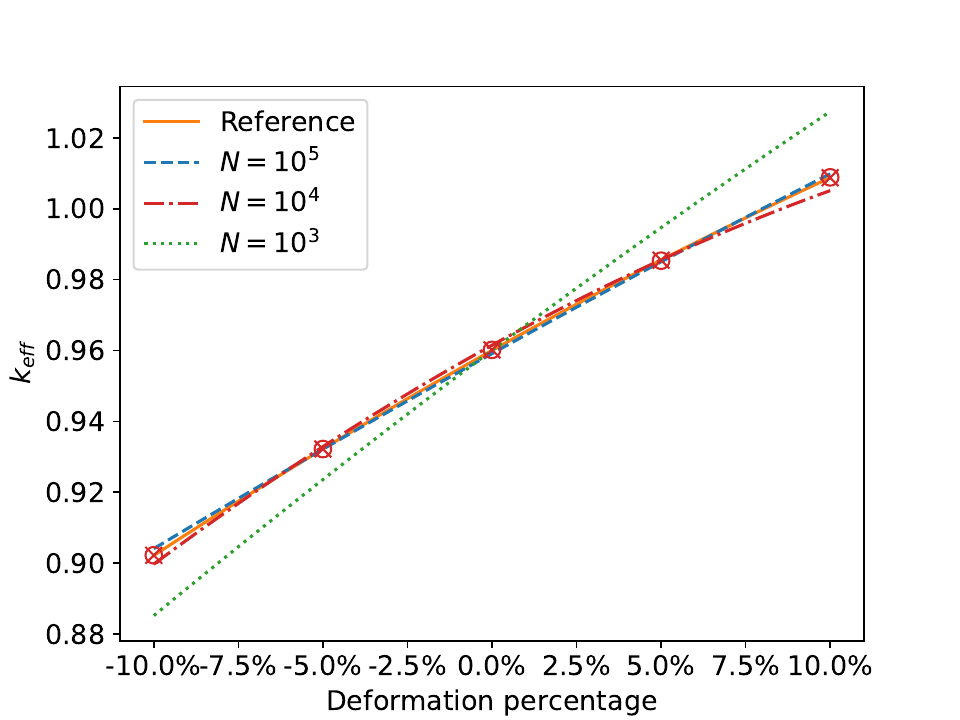}
        \caption{Convergence in initial population size.}
        \label{fig:uni_box_N}
    \end{subfigure}
    \caption{Left: Convergence of the gPC model in polynomial order for the uniform expansion of the box system. $P$ denotes the polynomial order. Right: convergence of the gPC model as a function of the initial population size $N_S$. The total number of histories $H = N_S \times G_a =10^8$ is preserved for all but the reference curve which is the same as in~\ref{fig:uni_imm_box_exp} and is obtained using $N_S=10^6$, $G_a=10^3$ and $H=10^9$ neutron histories. $P=2$ for all curves in the right plot.}
    \label{fig:uni_box_conv}
\end{figure}

In addition, the convergence of the gPC model with the polynomial order for the isotropic expansion is shown in Figure~\ref{fig:uni_box_conv_P}. The perturbation amplitude ranges from halving the size of the box to doubling it. It should be noted that at first order, the gPC model may considerably differ from reference results, even at $X=0$. This is because all even order Legendre polynomials contribute at $X=0$. This is in stark contrast with regular perturbation theory based on a Taylor expansion, which should always coincide with reference results at $X=0$. On the other hand, convergence is achieved very quickly even for so large a perturbation, as there is little difference between $P=3$ and $P=4$. This property of generalized polynomial chaos could be put to good use when one is interested in large range of reactivity, e.g. for critical searches. Moreover, future work should focus on a systematic comparison between the gPC approach and the perturbation theory for geometrical changes developed in~\cite{yamamoto_monte_2018} and~\cite{yamamoto_monte_2021}. 

Lastly, the convergence of the gPC model as a function of the initial population size is described in Figure~\ref{fig:uni_box_N} for the isotropic expansion of the homogeneous box. We recall that in our method, the initial population size also corresponds to the number of samples for the uncertain parameters, as each initial particle is assigned with a sample of the uncertain parameters that it transmits to its fission progeny. In this figure, it appears clearly that if the initial population size is too small (i.e. the initial number of random parameters does not adequately cover the space of available uncertain parameters), the gPC model fails to adequately reproduce the reference results. The reference results correspond to the gPC model used in Figure~\ref{fig:uni_box_exp}. For $N_S=10^3$, the reactivities predicted by the gPC model are clearly outside the values allowed by the statistical uncertainties. When $N_S=10^4$, the predictions of the gPC model still lies outside of statistical uncertainties compared to the reference results but are significantly closer, while for $N_S=10^5$ there is statistical agreements within $3\sigma$ between the reference results and the gPC predictions. The same qualitative behaviour was observed for both estimator $k_\text{eff,1}$ and $k_\text{eff,2}$, although we show the result for estimator $k_\text{eff,2}$ due to its better statistical accuracy. Therefore, when performing gPC calculations using our algorithm, increasing the population size within a generation is preferable over increasing the number of active generations, as it is in fact usually the case in Monte Carlo power iteration calculations. Populations that are too small can lead to a bias in the estimation of the polynomial chaos coefficients. Note that we have verified that the bias arising from lower population sizes was not caused by insufficient statistical convergence - all coefficients were converged within $1\%$ relative error.

\begin{figure}
    \centering
    \begin{subfigure}{0.45\textwidth}
        \includegraphics[width=\textwidth]{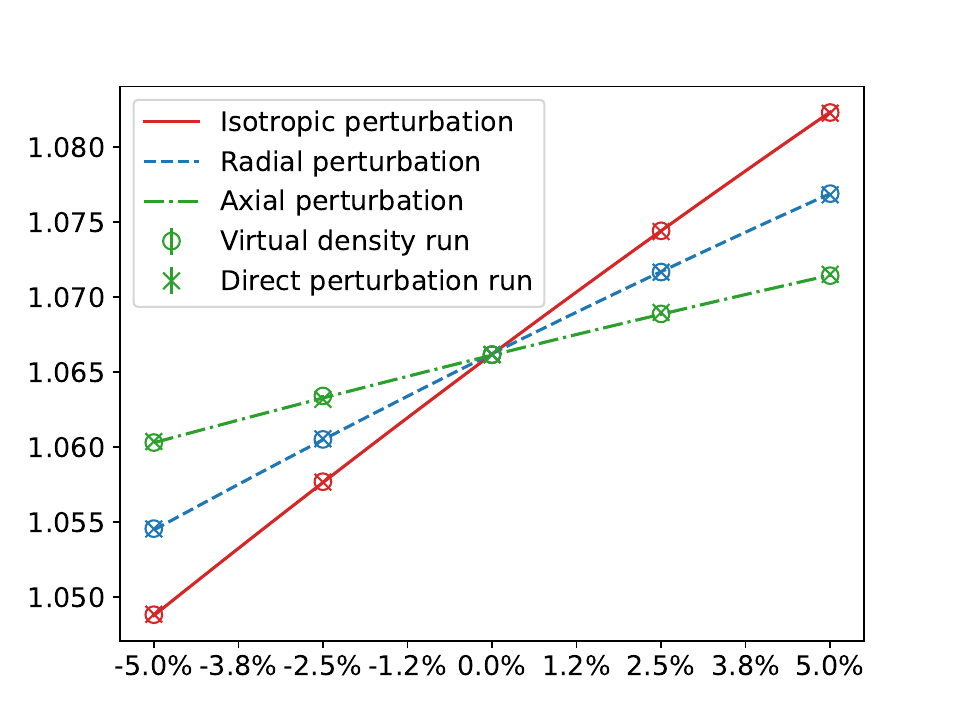}
        \caption{Uniform expansion}
        \label{fig:uni_imm_box_exp}
    \end{subfigure}
    \begin{subfigure}{0.45\textwidth}
        \includegraphics[width=\textwidth]{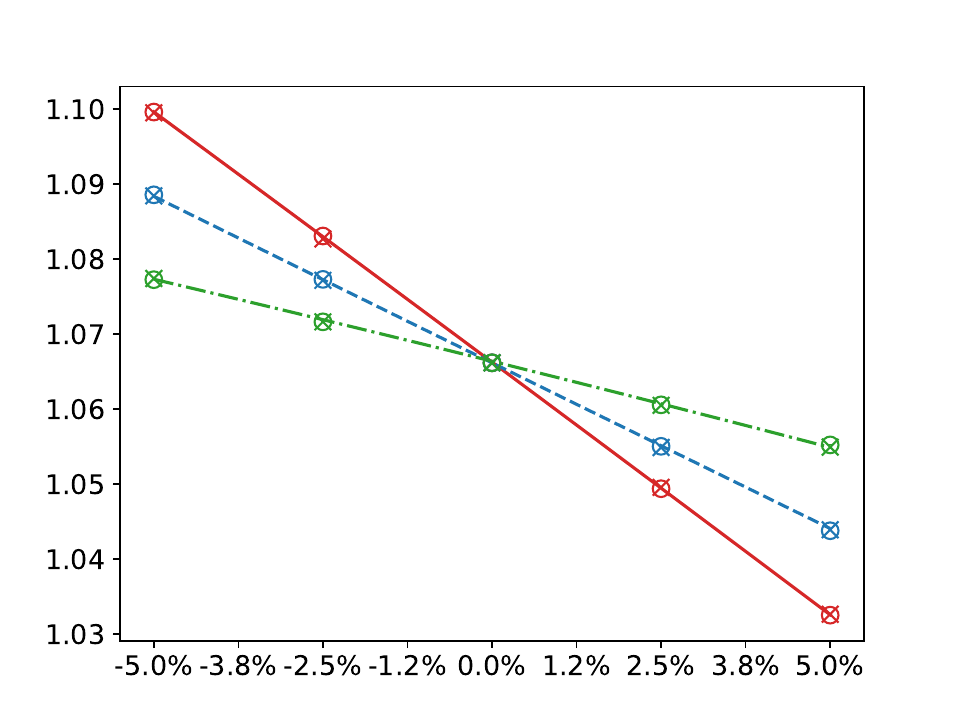}
        \caption{Uniform swelling}
        \label{fig:uni_imm_box_swl}
    \end{subfigure}
    \caption{Numerical comparison between virtual density, direct calculations, and gPC using virtual density for uniform swelling or expansion of the reflected system. Full lines represent the (continuous) gPC model, circles stand for virtual density and crosses for direct perturbations.}
    \label{fig:uni_imm_box}
\end{figure}

The system investigated previously is relatively simple. In particular, even if we were using regular correlated sampling (CS) or differential operator sampling (DOS), we would expect no significant indirect effect, i.e. no significant effect of the geometrical perturbation on the fission source~\cite{morillon_use_1998, nagaya_impact_2005}. Our second system is an idealization of a reflector problem, where it would be expected that, if our method were subjected to indirect effects, it would appear clearly. Note that there is no theoretical reason for any indirect effect to appear in our method: indirect effects in DOS and CS appear because the adjoint flux is replaced with the direct flux in formulas obtained through perturbation theory, and the effect of the perturbation on the fission source is therefore lost unless special care is taken~\cite{raskach_improvement_2009, yamamoto_exact_2021}. In our case, no such simplification is required. The effect of the perturbation is directly simulated, and its effect on the fission source is natively taken into account. $G_d=100$ inactive generations were sufficient to reach stationary state, and the gPC coefficients were obtained with $G_a =10^3$ active generations and $N_S=10^6$ neutrons. The results for the reflector case are shown in Figure~\ref{fig:uni_imm_box} for both swelling and expansion perturbations. Agreement between the gPC model and reference results is once again very good and within statistical uncertainties, and illustrates the fact that our methods should be free from indirect effects.

\subsection{Numerical results for the C5G7 assembly}

\begin{figure}
    \centering
    \begin{subfigure}{0.45\textwidth}
        \includegraphics[width=\textwidth]{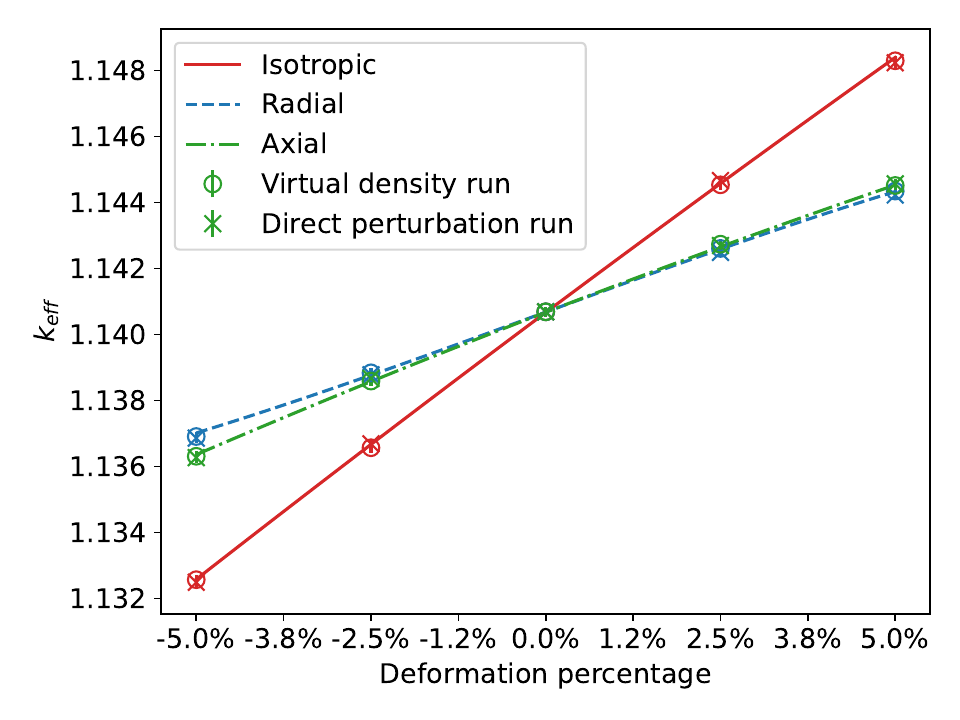}
        \caption{Uniform expansion}
        \label{fig:uni_c5g7_exp}
    \end{subfigure}
    \begin{subfigure}{0.45\textwidth}
        \includegraphics[width=\textwidth]{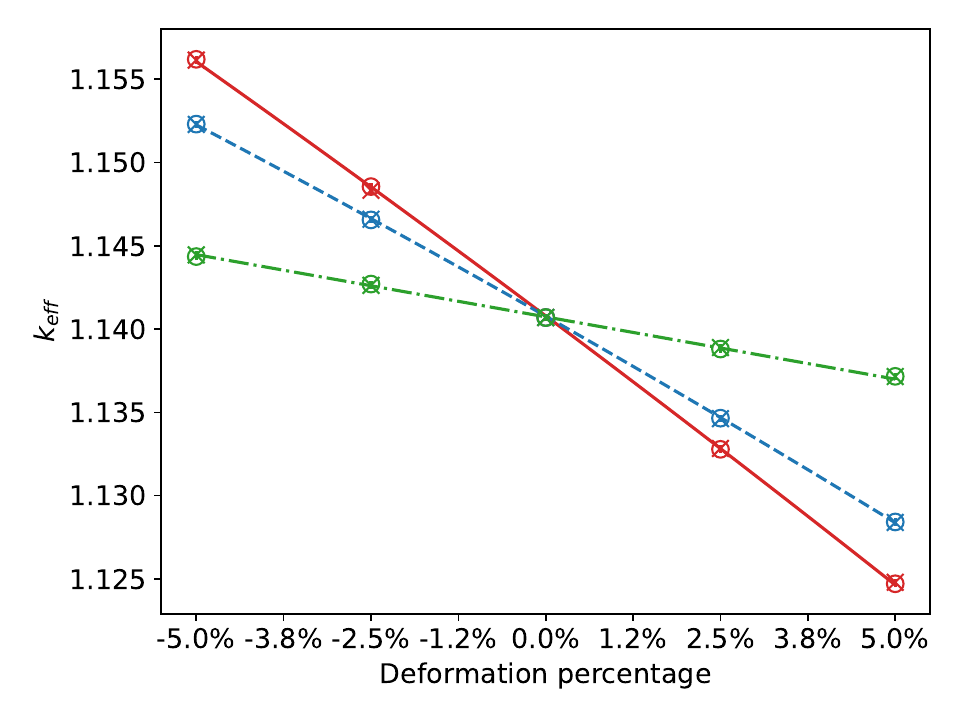}
        \caption{Uniform swelling}
        \label{fig:uni_c5g7_swl}
    \end{subfigure}
    \caption{Numerical comparison between virtual density or direct calculations, and gPC using virtual density for uniform expansion or swelling of the 3D-C5G7 assembly. Full lines represent the (continuous) gPC model, circles stand for virtual density and crosses for direct perturbations.}
    \label{fig:uni_assembly}
\end{figure}

The extent of reactivity changes due to geometrical perturbations investigated in the previous paragraph is considerably larger than what would be expected from thermal expansion of components in a realistic reactor core for example. More importantly than such large reactivity changes, the method should be able to accurately reproduce small changes in reactivity, which is better illustrated in Fig.~\ref{fig:uni_assembly} for uniform deformations of a 3D MOX assembly of the C5G7 benchmark. The reference virtual density and direct perturbation simulations were performed with $N_S=10^5$ neutrons and $G_a=1000$ active generations, while the gPC calculation required $N_S=10^7$ neutrons and $G_a=10^2$ active generations. In both cases, $G_d=100$ inactive cycles were sufficient to ensure convergence of the Shannon entropy. Additionally, a second order expansion was found to be sufficient to accurately reproduce Monte Carlo results for isotropic and axial perturbations, while a third order expansion was required for radial perturbations. All numerical results are once again found to be in very good agreement and within statistical uncertainty for either expansion and swelling. Radial and axial expansions are shown to have roughly the same impact on the multiplication factor, while the effect of isotropic expansions is roughly the sum of the effects of the former perturbations. This is significantly different from what is seen in Fig.~\ref{fig:uni_c5g7_swl}: there, radial swelling has a significantly more pronounced effect than axial swelling. This is naturally explained by the fact that axial swelling modifies leakage through the $x$ and $y$ boundaries, which are assigned with periodic boundary conditions, whereas radial swelling modifies leakage through the $z$ boundary, which is assigned with a proper leakage boundary condition. The effect of isotropic perturbations is once again roughly equal to the sum of that of the two other kind of perturbations. Overall, the effect of expansion on reactivity is smaller than that of swelling, which is a consistent trend over all of our results.  

Once again, we provide details on the accuracy of the methods in Table~\ref{tab:err_uni_exp_assembly} for the uniform, isotropic expansion and swelling of the assembly. Relative errors between virtual density and direct perturbation are always under about $3\%$. Moreover, it always lies within statistical uncertainty. The same remarks apply to the predictions of the gPC model. The trends are similar for axial and radial perturbations. While the relative error is larger than in Table~\ref{tab:err_uni_exp_box}, it is only because the magnitude of the reactivity changes is much smaller than for the previous systems.

\begin{table}[]
    \centering
    \begin{tabular}{ c c c c c c c }
    \hline
    {\bf Swelling} & $\Delta\rho_\text{direct}$ & $\Delta\rho_\text{virtual}$ & $\delta(\Delta\rho_\text{virtual})$ & $\Delta\rho_\text{gPC}$ & $\delta(\Delta\rho_\text{gPC})$ \\ \hline
    $-5\%$ & $-1171 \pm 12$ & $-1175 \pm 12$ & $-0.3\%$ & $-1164$ & $0.57\%$ \\ 
    $-2.5\%$ & $-583 \pm 12$ & $-599 \pm 12$ & $-2.6\%$ & $-598$ & $-2.9\%$ \\ 
    $2.5\%$ & $608 \pm 12$ & $612 \pm 12$ & $-0.6\%$ & $609$ & $-0.07\%$ \\ 
    $5\%$ & $ 1238\pm 12$ & $-1245 \pm 12$ & $-0.6\%$ & $-1251$ & $-1.0\%$ \\ 
    \hline
    {\bf Expansion} &  &  & &\\ \hline
    $-5\%$ & $635 \pm 12$ & $629 \pm 12$ & $1.0\%$ & $628$ & $1.1\%$ \\ 
    $-2.5\%$ & $308 \pm 12$ & $317 \pm 12$ & $-3.0\%$ & $310$ & $-0.5\%$ \\ 
    $2.5\%$ & $-304 \pm 12$ & $-294 \pm 12$ & $3.3\%$ & $-297$ & $2.3\%$ \\ 
    $5\%$ & $ -576\pm 12$ & $-581 \pm 12$ & $-0.8\%$ & $-589$ & $-2.1\%$ \\ 

    \hline
\end{tabular}
    \caption{Reactivity changes due to isotropic uniform perturbations of the C5G7 assembly. Reactivity perturbations are specified in $\text{pcm}$.}
    \label{tab:err_uni_exp_assembly} 
\end{table}

\begin{figure}
    \centering
    \includegraphics[width=0.45\textwidth]{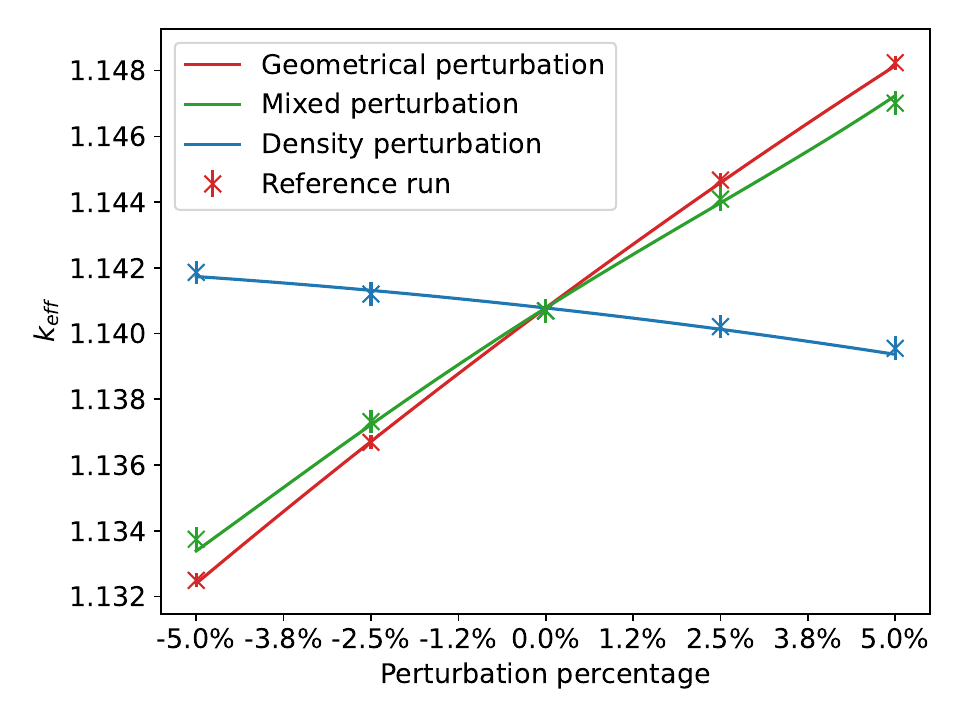}
    \caption{Numerical comparison between reference calculations, and gPC using virtual density for an isotropic uniform expansion and boron density perturbation of the 3D-C5G7 assembly. Crosses denote reference runs with $3\sigma$ uncertainty, and lines denote different evaluations of the gPC model.}
    \label{fig:uni_assembly_2D}
\end{figure}

Finally, for the sake of completeness and to demonstrate that our method is able to resolve multidimensional perturbations, the natural  boron density in the water is perturbed alongside an isotropic expansion of the 3D-C5G7 assembly, such that $\mathbf{X}=(X_1,X_2)^T$ is the uncertain vector, and $X_1,X_2\sim {\cal U}[-1,1]$ iid. $X_1$ parametrises the amplitude of the isotropic geometrical expansion, and the boron isotopic densities are perturbed such that
\begin{align}
    N_{^{10}B}(X_2) &= N_{^{10}N}^0 \times(1+X_2 \epsilon_2) \\
    N_{^{11}B}(X_2) &= N_{^{11}N}^0 \times(1+X_2 \epsilon_2),
\end{align}
where $\epsilon_2=0.05$ denotes the maximum perturbation amplitude and $N^0_{^{10}N}$ and $N^0_{^{11}N}$ denote the unperturbed boron isotopic concentrations as found in~\cite{nea_benchmark_2000}. Figure~\ref{fig:uni_assembly_2D} compares the predictions from the gPC model with reference calculations in which the geometry and nuclide densities are explicitly modified in the input file. The simulation parameters $N_S$, $G_a$ and $G_d$ for the gPC calculations were set to $N_S=5\times10^7$, $G_a=10^2$, and $G_d=20$. $G_d=20$ was chosen to minimize calculation time, and was verified to be sufficient to reach stationarity. As expected from the literature~\cite{poette_numerical_2022}, we have observed that higher order gPC coefficients become significantly noisier when the dimension increases, as illustrated by Table~\ref{tab:2d_gpcModel}, which prompted us to increase $N_S$ compared to the previous C5G7 calculations. Reference results were obtained with an average $1\sigma$ uncertainty of $10$ pcm. The gPC model was truncated to second order in each uncertain direction. With the computational power allocated to this calculation, third order coefficients were too noisy to be meaningful. As mentioned in Section~\ref{sec:combination}, for this case the polynomial approximation of $\nu^g$ was replaced by a 2D histogram approximation with $N_{X_1}=N_{X_2}=100$ bins. The red curve corresponds to the gPC model evaluated for $(X_1,0), X_1\in[-1,1]$, i.e. no density perturbation and the same isotropic expansion as in Figure~\ref{fig:uni_c5g7_exp}. The blue curve corresponds to the gPC model evaluated for $(0,X_2), X_2\in[-1,1]$, i.e. no geometrical perturbation and only a density perturbation on the boron isotopic densities. The green line corresponds to the evaluation of the gPC model for simultaneous geometrical and density perturbations of equal magnitude, denoted as the mixed perturbation. Figure~\ref{fig:uni_assembly_2D} shows that the predictions from the gPC model agree within $3\sigma$ uncertainty with the reference results. Inspection of Table~\ref{tab:2d_gpcModel}, in which the value of the gPC coefficients and the associated statistical uncertainty are presented, suggests that additional acceleration methods targeted at the uncertain dimension should be considered, such as in~\cite{poette_multigroup-like_2023}. Nevertheless, the method is able to accurately capture the effect of simultaneously perturbing the boron isotopic densities and the geometry. The 2D gPC reduced order model is obtained in a single calculation and yields the full dependency of $k_{eff}$ with respect to the amplitude of the two considered perturbations, including cross-terms, i.e., terms akin to mixed derivatives in regular perturbation theory. Those terms encode the corrections due to the fact that perturbing the geometry modifies the sensitivity of the system to density perturbations, and vice versa, and are typically hard to obtain using traditional perturbation theory for geometrical changes.

\begin{table}[]
    \centering
    \begin{tabular}{ c c c c }
    \hline
    $k^{k_1,k_2}_{eff}$ & $k_1=0$ & $k_1=1$ & $k_1=2$ \\ \hline
    $k_2=0$ & \num{1.14057} & \num{-1.18690e-3} & \num{-7.92879e-5} \\ 
    $k_2=1$ & \num{7.95594e-3} & \num{-6.06993e-5} & \num{1.61916e-4} \\ 
    $k_2=2$ & \num{-2.64954e-4} & \num{-6.37776e-6} & \num{-1.45610e-4} \\ 
    \hline
    \\
    \hline
    $\Delta k^{k_1,k_2}_{eff}$ & $k_1=0$ & $k_1=1$ & $k_1=2$ \\ \hline
    $k_2=0$ & \num{1.14057e-5} & \num{2.14992e-5} & \num{2.78344e-5} \\ 
    $k_2=1$ & \num{2.25895e-5} & \num{3.68465e-5} & \num{5.08528e-5} \\ 
    $k_2=2$ & \num{2.79034e-5} & \num{5.10183e-5} & \num{6.60395e-5} \\ 
    \hline
\end{tabular}
    \caption{Coefficients of the 2D gPC reduced order model associated with Figure~\ref{fig:uni_assembly_2D}. The upper table describes the value of the coefficients, where $k_1$ corresponds to the order of the geometrical perturbation and $k_2$ to the order of the density perturbation. The lower table gives the associated $1\sigma$ uncertainty.} 
    \label{tab:2d_gpcModel} 
\end{table}

\section{Conclusions}
\label{sec:conclusions}

In this work, we have introduced an efficient method for computing the effects of geometrical perturbations in a static $k$-eigenvalue Monte Carlo power iteration by combining the virtual density theory with the intrusive generalized polynomial chaos method. We have given a Monte Carlo algorithm that is adapted to $k$-eigenvalue simulations traditionally performed for reactor analysis and that can be incorporated in pre-existing codes with relatively minor changes. Because of our choice to use a unique Legendre polynomial basis, it is essential to ensure that the distribution of uncertain parameters in the fission source at the start of each generation remain uniform and stationary. The method can be extended in a straightforward manner to compute sensitivity coefficients for both geometrical and cross-section perturbations as well as to propagate uncertainties. While geometrical perturbation theory in Monte Carlo simulations is not straightforward to implement and requires careful book-keeping or knowledge of the adjoint flux, the coefficients of a reduced model based on polynomial chaos are easily estimated in a Monte Carlo simulation, and give access to the full dependency of $k_\text{eff}$ for a large range of perturbations with a single calculation. The developed algorithm converges quickly in polynomial order, although no systematic study has been done here regarding this specific property of the method. We have shown that using a polynomial chaos approach yields accurate results even for realistic geometries in a continuous energy framework such as a fuel assembly. Moreover, it appears to be free of indirect effects.

In order to improve the reliability of the method, future works should investigate in detail the effect of the population control algorithm through which the distribution of uncertain parameters at the start of a generation is made stationary, as well as more robust moment preserving techniques for approximating the distribution of uncertain parameters at the end of each generation, since the inverse of this distribution is utilised as an importance function for population control. Alternatively, one could avoid the need to use Legendre polynomials, and therefore population control, by allowing scoring on an arbitrary orthogonal polynomial basis, which would however vary with generations.

In the current work, our method was restricted to uniform perturbations, i.e., the same constant perturbation was applied to the entirety of the geometry. Future works will focus on extending our work to non-uniform perturbations. We should note that there is no theoretical issue with this extension, as it simply requires that an adequate coordinate change algorithm for direct simulations of non-uniform perturbations is implemented in our code. The rest of the method would be unchanged. Another venue for application would be in criticality searches, once non-uniform perturbations can be taken into account.  

A limitation of the traditional perturbation theory method for geometrical perturbations lies in its first order restriction when considering simultaneous perturbations of internal boundaries and nuclide densities. However, our method should be able to naturally estimate the contribution of higher order effects on the reactivity. This has been illustrated in the simple case of the perturbation of boron isotope densities alongside a uniform isotropic expansion of the 3D-C5G7 assembly.

Further works should focus on the behaviour of the algorithm as the number of perturbed parameters increases. In agreement with the literature, higher order coefficients become noisier as the dimension of the uncertain space increases, suggesting that special care should be taken to adequately minimize the noise in the estimation of higher order coefficients, for e.g., by implementing the method proposed in~\cite{poette_multigroup-like_2023}.

Investigating continuous or time-dependent geometrical changes using the virtual density theory would also be an interesting venue. Note that in this work, the virtual density theory the was used to explicitly model the geometrical perturbations instead of being used to assess the corresponding sensitivity coefficients of the system. Therefore, it would be interesting to compare our results and the efficiency of our method with more traditional perturbation theory, as developed in~\cite{yamamoto_monte_2021, yamamoto_monte_2025}.

\section*{Acknowledgement}

This work was partially supported by the UK EPSRC program grant Mathematical Theory of Radiation Transport: Nuclear Technology Frontiers (MaThRad), Grant number EP/W026899/2. We would also like to thank Mikolaj Adam Kowalski for his invaluable assistance with implementation of geometrical perturbations through coordinate changes, and to Valeria Raffuzzi for insightful discussions. 

\appendix

\section{Importance combing on the uncertain dimension}
\label{appendix1}
The importance-based comb on the uncertain dimension presented in Algorithm~\ref{algo:alg1} can be otherwise written in a more mathematical form. To that end, we redefine some necessary quantities.

Let us denote by $\mathbf{n}_F$ the set of particles to comb, and by $\mathbf{n}_S$ the set of selected particles, after combing. A particle $n_i\in\mathbf{n}_F,\mathbf{n}_S$ is assigned with a coordinate vector in the extended phase space, such that we denote 
\begin{equation}
    n_i=(\mathbf{r}_i,\mathbf{\Omega}_i, E_i, \mathbf{X}_i,w_i),
\end{equation}
with $\mathbf{X}_i\in V$ the vector of uncertain parameters, and $N_F=card(\mathbf{n}_F)$, where $card(A)$ denotes the number of elements in a set $A$. In what follows, we assume unit statistical weight and suppress $w_i$, as this is sufficient for our purpose and simplifies the notations. The same realization of $\mathbf{X}$ is typically shared by several particles, and we define the set of distinct uncertain vectors
\begin{equation}
    {\cal{Q}} = \{\mathbf{X_i}:n_i\in\mathbf{n}_F\},
\end{equation}
and the total number of distinct uncertain vectors in $\mathbf{n}_F$, $\tilde{N}=card({\cal Q})$. We then define the multiplicity of $\mathbf{X}\in{\cal Q}$ by
\begin{equation}
    \alpha_k=card(\{n_i\in\mathbf{n}_F:\mathbf{X}_i=\mathbf{X}_k\}), \mathbf{X}_k\in{\cal Q}.
\end{equation}
Then we define 
\begin{align}
    \nu(\mathbf{X_i})&=\mathbb{P}(\mathbf{X}_i|n_i\in \mathbf{n}_F) =\frac{\alpha_i}{N_F}
\end{align}
and
\begin{align}
    \mu(\mathbf{X}_i)=\mathbb{P}(\mathbf{X}_i|n_i\in\mathbf{n}_S).
\end{align}
We want to select $N_S$ particles from $\mathbf{n}_F$ such that $\mu(\mathbf{X}_i)=K\in\mathbb{R}$, i.e. such that the probability distribution of uncertain vectors after combing is uniform. This can be done by adapting the importance combing presented in~\cite{thomas_e_booth_weight_1996} in the following way, where $i$ denotes the index of a particle in $\mathbf{n}_F$. We define
\begin{align}
    I_i&=\frac{1}{\nu(\mathbf{X}_i)}=\frac{N_F}{\alpha_i},\\
    I&=\sum_{i=1}^{N_F} I_i=N_F\tilde{N}.
\end{align}
Define the importance weighted comb by
\begin{align}
    t_m&=\xi \frac{I}{N_S}+(m-1)\frac{I}{N_S} \text{ and } \xi\sim{\cal U}[0,1], m\in\{1,\dots,N_S\},
\end{align}
where $\xi$ is simply an initial offset, sampled only once. Then,
\begin{equation}
    j < \frac{I_i}{\frac{I}{N_S}} < j+1
\end{equation}
with $j$ an integer. The number of times $J$ the particle $i$ is copied is $j$ with probability
\begin{equation}
    p_{i,j} = j + 1 - \frac{I_i}{I/N_S},
\end{equation}
and $j+1$ with probability
\begin{equation}
    p_{i,j+1} = \frac{I_i}{I/N_S}-j.
\end{equation}

Each time a particle is copied, it is assigned a new statistical weight $w'_i$ that needs to be determined. However, $J$ is not the number of times $\mathbf{X}_i$ is selected, as $\mathbf{X}_i$ can be shared by several particles in $\mathbf{n}_F$. The expected number of times the particle $i$ is selected is $\frac{N_S}{\tilde{N}\alpha_i}$. Summing over all particles in $\mathbf{n}_F$ that share $\mathbf{X}_i$, we get that the expected number of times $\mathbf{X}_i$ is selected is given by:
\begin{equation}
    N_{\mathbf{X}_i}= \frac{N_S}{\tilde{N}}\sum_{k=1}^{N_F} \frac{\delta_{\mathbf{X_i},\mathbf{X_k}}}{\alpha_i}=\frac{N_S}{\tilde{N}},
\end{equation}
where $\delta_{a,b}$ is the usual Kronecker symbol. Remembering that selected particles are added to $\mathbf{n}_S$, it follows that setting $w'_i=1$ for all $n_i\in\mathbf{n}_S$, implies by construction and after normalization that
\begin{equation}
    \mathbb{E}[\mu(\mathbf{X}_i)]=\mathbb{E}[\mathbb{P}(\mathbf{X}_i|n_i\in\mathbf{n}_S)]=\frac{1}{\tilde{N}},
\end{equation}
i.e. the uncertain vectors carried by particles in $\mathbf{n}_S$ are uniformly distributed \textit{on average}. Note that contrary to the algorithm presented in~\cite{thomas_e_booth_weight_1996}, this algorithm is not weight-preserving, nor does it need to be. In the generation context in which Algorithm~\ref{algo:alg1} is introduced, $\mathbf{n}_S,\mathbf{n}_F,$ and ${\cal Q}$ need to be replaced by their generation-wise equivalent.
\bibliographystyle{alpha}
\bibliography{references}

\end{document}